\documentclass{vgtc}                          

\graphicspath{{figures/}{pictures/}{images/}{./}} 

\usepackage{times}                     

\usepackage{tabu}                      
\usepackage{booktabs}                  
\usepackage{lipsum}                    
\usepackage{mwe}                       

\usepackage{tipa}

\usepackage{mathptmx}                  

\onlineid{1009}

\vgtccategory{Research}

\vgtcinsertpkg

\usepackage{enumitem}
\usepackage{fontawesome}
\usepackage{colortbl}
\usepackage{url}
\usepackage{float}
\usepackage{pifont}
\usepackage{multirow}
\usepackage{listings}
\usepackage{tcolorbox}

\newcommand{\pheading}[1]{\vspace{2px}\noindent\textbf{#1}}
\newcolumntype{R}[1]{>{\raggedright\arraybackslash}p{#1}}

\arrayrulecolor{gray} 

\newenvironment{tightItemize}{\begin{itemize} \itemsep
-2.1pt}{\end{itemize}}

\usepackage{tabularx}
\usepackage{tabu}                      
\newcolumntype{P}[1]{>{\arraybackslash}p{#1}}

\usepackage{xurl}

\usepackage{booktabs}
\usepackage{color}[svgnames] 
\definecolor{codegreen}{rgb}{0,0.6,0}
\definecolor{codegray}{rgb}{0.5,0.5,0.5}
\definecolor{codepurple}{rgb}{0.58,0,0.82}
\definecolor{backcolour}{rgb}{0.894, 0.909, 0.925}

\title{Voicing Uncertainty: How Speech, Text, and Visualizations Influence Decisions with Data Uncertainty}

\author{Chase Stokes\thanks{e-mail: cstokes@ischool.berkeley.edu}\\ %
        \scriptsize University of California Berkeley
    \and Chelsea Sanker\thanks{email:sanker@stanford.edu}\\ %
        \scriptsize Stanford University 
    \and Bridget Cogley\thanks{email:bcogley@versalytix.com}\\ %
        \scriptsize Versalytix
    \and Vidya Setlur\thanks{e-mail: vsetlur@tableau.com}\\ %
        \scriptsize Tableau Research}

\abstract{
Understanding and communicating data uncertainty is crucial for informed decision-making across various domains, including finance, healthcare, and public policy. This study investigates the impact of gender and acoustic variables on decision-making, confidence, and trust through a crowdsourced experiment. We compared visualization-only representations of uncertainty to text-forward and speech-forward bimodal representations, including multiple synthetic voices across gender. Speech-forward representations led to an increase in risky decisions, and text-forward representations led to lower confidence. Contrary to prior work, speech-forward forecasts did not receive higher ratings of trust. Higher normalized pitch led to a slight increase in decision confidence, but other voice characteristics had minimal impact on decisions and trust. An exploratory analysis of accented speech showed consistent results with the main experiment and additionally indicated lower trust ratings for information presented in Indian and Kenyan accents. The results underscore the importance of considering acoustic and contextual factors in presentation of data uncertainty.
} 


\begin{document}



\maketitle

\section{Introduction} 
\label{sec:intro}

In today's world, where decisions are increasingly data-driven, effectively conveying uncertainty is critical for making sound choices across domains such as finance, healthcare, and public policy~\cite{draper:1995, Morss2008CommunicatingUI}. Data uncertainty encompasses a range of potential outcomes, variability within datasets, and possible errors in  predictions~\cite{Morgan1990UncertaintyAG}. 
While precise data would ideally drive decision-making, exact information is often unavailable in real-world scenarios. Thus, communicating uncertainty becomes essential to understanding the true state of the data. 
A primary challenge in communicating data uncertainty lies in its interpretation. This task is fraught with challenges around trust, reliability, and bias~\cite{CapurroEtAl2021, hullman2020why, sacha:2016,tversky1974judgment}. While experts might grasp statistical nuances like confidence intervals or p-values, these concepts tend to be confusing for a lay audience~\cite{Spiegelhalter:20111}. 

Traditional methods for conveying uncertainty have predominantly focused on visual and textual tools. Visual aids such as error bars, confidence intervals, and density plots help illustrate data variability, scope, and distribution~\cite{padillabook:2021}. In text, hedge words like ``somewhat" and ``possibly" signal uncertainty~\cite{Lakoff1973}. 
In considering spoken communication, features such as pitch and speech rate can indicate a speaker's uncertainty or hesitation~\cite{BrennanWilliams1995, SchererEtAl1973}. 
However, each mode of communication has its trade-offs. Visualizations require a level of graphical literacy that not all viewers possess, while lengthy textual explanations can lead to limited comprehension. The transient nature of speech restricts the ability to revisit information compared to text or visualizations.

Recent research has highlighted the potential of multimodal approaches of visualizations, text, and speech to improve the communication of uncertainty~\cite{stokes2024delays,stokes2024mixing}. Speech, in particular, is relevant for scenarios where quick, informed decisions need to be made that rely on visual data that may be verbally communicated with inherent uncertainties, such as in voice-activated systems (e.g., a voice assistant communicating uncertain weather forecasts), telemedicine (e.g., doctors or AI systems communicating uncertainty about diagnoses or treatment outcomes), and public policy announcements (e.g., officials communicating complex and uncertain information during a public health crisis). Despite the potential richness of speech in conveying nuanced information, the role of acoustic variations in communicating uncertainty remains an understudied area; existing research has largely overlooked how different speech parameters, such as pitch, speech rate, and speaker characteristics like gender\footnote{We employ the terms ``Woman" and ``Man" to align with contemporary norms which emphasize social and cultural roles of gender identity rather than biological attributes. We acknowledge that these categories are not representative of all genders \cite{Thorne2019TheTO}. However, synthesized speech sources such as Microsoft Azure only offer binary options. As such, we move forward in the paper, focusing exclusively on these two genders.} or accent affect the perception of uncertainty. Understanding the variation across different voices and vocal characteristics is important for comparing speech to other modes; some of the previously observed unique results for speech conditions could reflect something about the particular audio stimuli rather than a broader pattern of how people process auditory information. 

This work seeks to address these gaps by systematically investigating how variations in pitch, speaking rate, and speaker characteristics impact the use of data uncertainty. 
Through a crowdsourced experiment, we aim to further clarify and inform the trade-offs between different speech modalities and offer insights for designing effective multimodal uncertainty communication strategies. Specifically, our research contributions are: 
\vspace{-1mm}
\begin{tightItemize}
\item \textbf{Replicating and extending prior work on communication with different modes of information.} Our findings corroborate increased risk for speech representations and lower confidence for text representations but do not observe higher trust for speech representations. 
\item \textbf{Providing empirical evidence on the effects of gender and acoustic variables in speech on the use of uncertain data.} Results demonstrate that while gender and acoustic features among American voices do not significantly impact decision-making outcomes, the normalized pitch does have an impact on decision confidence.
\item \textbf{Exploratory investigations around accented speech and decision-making.} We completed an exploratory analysis of additional speech variants, including British, Kenyan, and Indian accents.
Preliminary findings suggest that Kenyan and Indian accents may receive lower ratings of trust compared to British and American accents. These results are exploratory and warrant further investigation.
\end{tightItemize}

\section{Related Work}
\label{sec:rw}

We examine prior work across three modes of data communication.

\subsection{Visualizing Uncertainty}
Research on visualizing uncertainty has developed a broad array of methods to assist in understanding and communicating data variability~\cite{liu:2016,padillaensemble:2017, Thomson2005ATF}. Showing uncertainty in data provides a more accurate representation of the underlying data and its limitations, but uncertainty can be challenging for readers to interpret correctly \cite{hullman2020why}.
Techniques such as error bars, confidence intervals, and density plots are commonly used to illustrate data variability and provide a graphical representation of potential outcomes~\cite{padillabook:2021}. However, these methods often require a level of graphical literacy that not all users possess and can add visual artifacts to the design \cite{correll2014error}, potentially leading to misinterpretations. 

Prior research indicates that density plots and quantile dot plots are effective ways to communicate data uncertainty \cite{fernandes2018uncertainty, kay2016ish, padilla2021uncertain}, but they can be complex to interpret if readers are unfamiliar with the encodings \cite{stokes2024delays}. In addition, research by Padilla et al.~\cite{padillabook:2021} and Franconeri et al.~\cite{franconeri2021science} highlights a more comprehensive spectrum of uncertainty visualization techniques and their empirical impacts on decision-making. For example, ensemble visualization, which involves displaying multiple potential outcomes or data scenarios, can help users better grasp the range of possibilities inherent in uncertain data. However, these methods also demand a higher cognitive load and may not be suitable for all audiences. In contrast, icon arrays can be effective for decision-making, particularly for viewers with low numeracy or working memory capacity \cite{bancilhon2023combining, galesic2009using}. Showing uncertainty with \textit{simple} visual techniques can lower the cognitive load of interpretation \cite{bancilhon2023combining}.

Depending on the complexity of the visual and the task, visualizations may need to be supplemented with speech or text to provide additional information and explanation. Visualizations can make it easier for readers to identify critical information, but the information itself may be more understandable in text \cite{ottley2019curious}. However, lengthy text might be distracting or cumbersome and reduce overall comprehension. This workshop paper builds on prior visualization research by comparing the effectiveness of visual modes against various speech variants to further enhance understanding of uncertainty communication.

\subsection{Text Representations of Uncertainty}
Lexical methods for conveying uncertainty typically involve the use of hedge words and phrases that indicate varying degrees of certainty, such as ``somewhat," ``possibly," or ``sort of" \cite{Lakoff1973, SzarvasEtAl2012}. These linguistic cues help signal the probabilistic nature of information and guide readers in interpreting data variability. 
The audience can also form impressions based on other characteristics of how a message is presented.  
For example, Hu and Pan~\cite{HuPan2024} find that users are more forgiving of AI service failures when the error is reported in an informal `cute' style than when it is reported in a formal style.

There is a trade-off between presenting information in text versus in speech audio. Some studies find that listener recall is higher for the information presented in text than for the same information presented in audio \cite{FurnhamGunterGreen1990, Sundar2000}. However, naturally produced speech does not typically have the same structure as a written text, which is an important consideration when producing texts to be read aloud. The mode of information might interact with its length; listeners remember more information from a short paragraph when it is presented auditorily rather than in text, but for a longer passage, reading the text results in better recall \cite{KintschEtAl1975}.

Prior research has shown that while text can effectively communicate uncertainty, it may reduce confidence in decision-making~\cite{stokes2024delays}. Presenting information via speech rather than text can increase the perceived trustworthiness~\cite{stokes2024delays, WaytzHeafnerEpley2014}. However, the difference between text and speech depends on context; Sundar~\cite{Sundar2000} finds that participants rate a text-only news site as more credible than an audio-only site. One of the limitations of text is that it lacks informative cues that are present in speech, such as tone or pitch.

\subsection{Speech Communication}
Speech offers a rich, multifaceted mode of communication that can convey uncertainty through acoustic features such as pitch, speech rate, and pauses. 
Despite its potential, the role of speech in communicating uncertainty has been relatively understudied. Existing research has typically focused on either visual or textual representations, with limited exploration of how acoustic variables in speech affect listener interpretation. 
Work examining how different voices are perceived usually focuses on direct evaluations of the voice rather than testing how the voice impacts how listeners interpret the information provided in that voice. 

Effects of gender on evaluations of speaker characteristics are variable. Some studies find that women's voices receive higher trust ratings~\cite{GoodmanMayhorn2023}, while others find the inverse \cite{LeeNassBrave2000}. 
The listener's gender can also play a role; listeners are more likely to trust a voice matching their own gender \cite{LeeNassBrave2000}. Men's voices tend to receive higher ratings of being authoritative \cite{TolmeijerEtAl2021} and competent \cite{AhnKimSung2022}, while women's voices receive higher ratings for a range of positive social traits such as being friendly, sincere \cite{TolmeijerEtAl2021}, warm \cite{AhnKimSung2022}, empathetic, and understanding \cite{MooshammerEtzrodt2022}. 
The impact of voice gender 
varies based on the task; listeners prefer women's voices for social tasks and men's voices for informational tasks \cite{LeeRatanPark2019}. 

Some specific acoustic characteristics have been demonstrated to impact the perception of speaker. 
One of the frequently studied variables is pitch (the rate of vibration of the vocal folds, more precisely called ``fundamental frequency (F0)''); pitch is usually normalized in order to compare it to the average pitch among voices of the same gender. 
Lower pitch increases perceived competence, trustworthiness \cite{OleszkiewiczEtAl2017}, and authoritativeness \cite{TolmeijerEtAl2021}.  On the other hand, voices with higher pitch are rated as being more cooperative \cite{KnowlesLittle2016} and friendlier \cite{TolmeijerEtAl2021}.

Listeners perceive utterances with faster speech rates as being more credible than slower utterances and are more likely to be persuaded by faster speech \cite{MehrabianWilliams1969, MillerEtAl1976}. Faster speech is also rated as more intelligent and more confident \cite{BrownGilesThakerar1985}. 
However, the positive views of faster speech may be mediated by the listener's own speech rate; Feldstein et al.~\cite{FeldsteinDohmCrown2001} find that listeners give the highest competence ratings to speakers whose speech rates are similar to their own. 

A few studies look at how the acoustic characteristics of an interlocutor or virtual assistant's voice impact decision-making. 
Pias et al.~\cite{pias2024impact} created an approximation of age and emotion differences based on manipulating pitch and speech rate; among women's voices, the combination of slower rate and higher pitch was found to be more persuasive, whereas, for men's voices, the combination of faster rate and lower pitch was found to be more persuasive.
However, Knight et al.~\cite{knight:2021} found that voice variants did not significantly impact investment decisions in an economic game. Instead, participants responded to their partner's behavior. 
These results might suggest that the effects of voice characteristics are outweighed when behavioral information is available.

Listeners' associations and biases about particular social groups can be extended to the linguistic characteristics produced by people in those groups. For example, speakers of African American English are perceived as less credible than speakers of General American English \cite{KurinecWeaver2019}, and speakers of British Received Pronunciation are perceived as more intelligent and more confident than speakers of other dialects \cite{Giles1971}. Negative evaluations are apparent across a range of non-standard accents (see \cite{FuertesEtAl2012} for a review).

This workshop paper builds on these findings by investigating how different acoustic features and speaker characteristics influence the perception of uncertainty.

\section{Study}
\label{sec:study}

\subsection{Motivation}
This study builds on prior work comparing
speech, text, and visual representations of uncertainty \cite{stokes2024delays, stokes2024mixing}. The work from Stokes et al. found that speech led to higher ratings of trust than visual or text information, though at times encouraged riskier decisions. That work only tested unimodal representations, and the speech conditions were limited to using one voice. We aimed to determine how generalizable these results are over different voices.

In multimodal representations of information, visualization often serves as the primary source of information, with text and speech providing guiding and supporting roles \cite{stokes2024mixing}. However, there are instances where text or speech may be more effective as the primary representation. Audiences may have visual impairments or reading difficulties. They may be consuming information in environments where visual attention is limited, such as while walking or driving. Furthermore, with the rise of voice-activated assistants and smart devices, understanding the effectiveness of speech-forward data communication is increasingly important. 

\subsection{Stimuli}
\label{sec:stimuli}

\begin{figure}[ht]
    \centering
    \includegraphics[width = 0.75\linewidth]{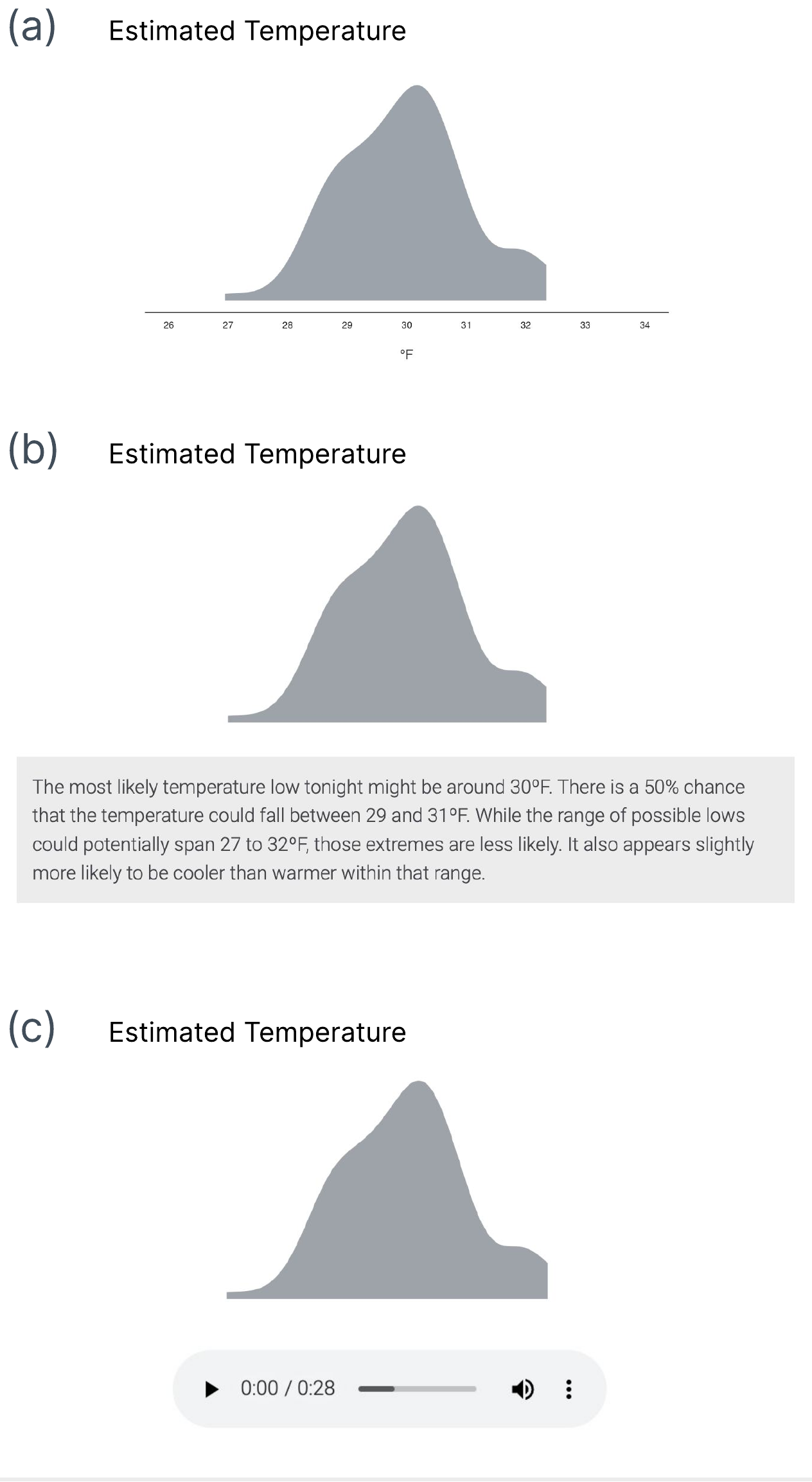}
    \caption{Example stimuli viewed by participants. (a) Visualization-only representation. (b) Text-forward representation. The complete text template can be found in \cref{sec:text_template}. (c) Speech-forward representation. Descriptions and links for speech-forward conditions can be found in \cref{sec:speech_links}. The example in this image can be found \href{https://osf.io/kxqfw}{here}.}  
    \label{fig:stimuli_examples}
\end{figure}

In this study, we examined bimodal representations of information, combining visualization with text or speech information. 
We compared eight different conditions in this study: one visual-only condition, one text-forward condition, and six speech-forward conditions. Examples of data presentations are shown in \cref{fig:stimuli_examples}. All supplemental materials, including stimuli, are available on OSF (\href{https://osf.io/mdz2y/}{link}), released under a CC BY 4.0 license.

In order to ensure that participants primarily used text or speech information to make their decisions, we provided them the visualization marks but removed the axes, as shown in \cref{fig:stimuli_examples}.
While this is not typical or recommended visualization practice, this representation isolates that \textit{visual} from the \textit{visualization}. 
By removing the x-axis, we still provided some visual information and context but could better isolate the impact of communicating information primarily through text or speech. In particular, we tested different speech variants (three women and three men). 

\subsubsection{Visual-only condition}
To create the distributions for the \textit{visual-only} condition, we generated a dataset with 100 data points by a normal distribution with a standard deviation of 1, created with the \texttt{rnorm} function in R (v4.3.0) \cite{rComputing}.
This method does not create perfect normal distributions but rather allows for a more ecologically valid set of stimuli since many distributions in the wild are not precisely normally distributed. The visual-only condition was a density plot \cite{ fernandes2018uncertainty, kay2016ish, stokes2024delays} created using the \texttt{ggdist} package in R \cite{ggdist}. 

\subsubsection{Text-forward condition}
\label{sec:text_template}
The \textit{text-forward} condition consisted of this same density plot with the x-axis removed. Below the visual mark, a paragraph described the distribution, including the most likely temperature, the middle quantile, the full range, and the skew of the distribution \cite{moments}. These features were extracted from the datasets generated for density plots.
The template used for text paragraphs was as follows:
\vspace{-4mm}
\begin{tcolorbox}[colback=gray!10, colframe=gray!10, boxrule=0pt]
The most likely temperature low tonight might be around [mean]ºF. There is a 50\% chance that the temperature could fall between [25th quantile] and [75th quantile]ºF. While the range of possible lows could potentially span [minimum] to [maximum]ºF, those extremes are less likely. It also appears [likelihood term] more likely to be [skew direction] within that range.
\end{tcolorbox}

\subsubsection{Speech-forward condition}
\label{sec:speech_links}
The \textit{speech-forward} conditions followed the same design. Below the visual mark, we positioned an mp3 player that contained the same information as present in the text paragraph. We applied the same adjustments as in Stokes et al. \cite{stokes2024delays}: a 0.2s delay, 5\% pitch decrease, and 70\% speed on numerical values, 5\% pitch decrease, and 65\% speed on hedge or likelihood terms. 

The speech stimuli used in this condition were generated using Microsoft Azure's Text-to-Speech (TTS) service~\cite{MicrosoftAzureTTS} in conjunction with Speech Synthesis Markup Language (SSML)~\cite{W3CSSML}. SSML allows for control over speech parameters such as pitch, rate, volume, and pauses. 
We use prosody and break times in the SSML to communicate uncertainty about temperature forecasts. Prosody adjustments include reducing the speech rate and slightly lowering the pitch of words providing key information. Break times, both short and longer pauses, are used to separate and highlight speculative elements of the message, such as hedge words like `could potentially' and `more likely.' Here is an example SSML snippet:

\begin{lstlisting}[language=XML]
<speak xmlns='http://www.w3.org/2001/10/synthesis' xml:lang='en-US'>
    <voice name='en-US-AmberNeural'>
       The <prosody rate='-35%' pitch='-5%'>most</prosody> likely temperature low tonight <prosody rate='-35%' pitch='-5%'>might</prosody> be <prosody rate='-35%' pitch='-5%'>around</prosody> <break time='0.1s'/><prosody rate='-30%' pitch='-5%'>thirty four</prosody> degrees Fahrenheit. 
    </voice>
</speak>
\end{lstlisting}

We sourced six voices (three labeled by Microsoft Azure as `female' and three as `male') to generate the snippets. These voices were selected based on their individual acoustic variations to ensure variation across conditions. Specifically, we chose the following voices with the corresponding descriptions provided by the Microsoft Speech Service Voice Gallery~\cite{MicrosoftSpeechService}. Example forecasts for each voice are linked below: 
\vspace{-2mm}
\begin{tightItemize}
\item Amber (\href{https://osf.io/rkwpy}{\textsc{Woman High}}): An engaging voice for children's stories that's warm and approachable, perfect for capturing the attention of young listeners. 
\item Ava (\href{https://osf.io/ubr9d}{\textsc{Woman Medium}}): A bright, engaging voice with a beautiful tone that's perfect for delivering search results and capturing users' attention.
\item Jane (\href{https://osf.io/bjdsr}{\textsc{Woman Low}}): An early-20s female voice like the girl next door that's warm and friendly, great for building a connection with users.
\item Guy (\href{https://osf.io/w9sp4}{\textsc{Man High}}): A friendly voice with slightly whimsical undertones and a wide expressive range that can convey any emotion with ease.
\item Davis (\href{https://osf.io/bv852}{\textsc{Man Medium}}): A generally calm and relaxed voice that can switch between tones seamlessly and be highly expressive when needed.
\item Eric (\href{https://osf.io/uwcs5}{\textsc{Man Low}}): A friendly voice that conveys soft-spoken confidence, inspiring trust and reliability with a calm and collected tone.
\end{tightItemize}

\subsection{Participants}

We used the G*Power software \cite{faul2009statistical, faul2007g} for power analysis, aiming to achieve a power of $0.95$ with an alpha threshold of $0.05$. Through post-hoc power analyses from Stokes et al. \cite{stokes2024delays}, we found a lower-bound effect size of $0.15$. Based on the power analysis for linear multiple regression with a maximum of 18 predictors (forecasts, speaker genders, participant genders, preferences, and speech features), the optimal sample size was 212 participants.

275 participants were recruited from Prolific \cite{palan2018prolific}. Participants were required to be located in the United States, have at least a 95\% acceptance rate, and be fluent in English. This population is meant to represent a sample of the general population, although subject to the demographic distributions present among Prolific users. They completed an $18$-minute survey and were compensated \$3.60. After excluding responses that did not pass our attention checks (n = 57), 218 participant responses were analyzed. 

Of these participants, 126 were women, 82 were men, 8 were non-binary, and 2 did not indicate their gender.
Participants were fairly educated on average, with 83 having a 4-year degree and 45 with at least some college. Only 2 participants had less than a high school education, and 34 were high school graduates. 
They also tended to be mostly young adults; 43 participants were 18-24 years old, and 74 were 25-34 years old. 
Participants had a moderate amount of experience with snow and ice. 
Most participants ($n = 127$) had lived more than 10 years in an area that received snow or ice during at least part of the year, and 64 participants had been responsible for applying salt to an icy road or walkway. Further details on demographics can be found in supplementary materials. 

\subsection{Method}

\subsubsection{Decision Framework}

We use a similar decision framework as prior work \cite{joslyn2012uncertainty, nadav2009uncertainty, padilla2021uncertain} and replicate the structure exactly from the work by Stokes et al. \cite{stokes2024delays}. Participants were asked to decide whether to apply salt to roads based on a low-temperature forecast for a given evening. In regions prone to icy conditions, applying salt to roads is a common practice to prevent accidents caused by slippery surfaces. Salt lowers the freezing point of water, helping to melt the ice that forms on road surfaces and reducing the likelihood of ice reformation.

They started with a fictional budget of \$12,000. Applying salt to the roads cost \$1,000. If the temperature of the given evening fell below 32ºF and the participant had not chosen to apply salt to the road, they were penalized \$3,000. Importantly, participants did not receive information as to the outcome of their choice, so there was no learning effect over the course of the survey. With this ratio of Cost:Penalty, the objective, rational choice was to apply salt to the roads if and only if the likelihood of the temperature falling below 32ºF was at or above 33\%. As an incentive to make cost-effective decisions, participants received a \$0.05 bonus for every \$1,000 remaining in the budget at the end of the survey.

\subsubsection{Survey Design}
Participants completed a Qualtrics \cite{qualtrics} survey with five main sections. They began the survey with an introduction to the task at hand, including a detailed explanation of the objectives. 
After being introduced to the task, they also received a description of the forecast they would use to make their decisions, including a description of how to interpret or use a density plot. This was displayed based on the condition assigned using embedded data in Qualtrics.
This was a between-participants design, meaning that each participant only viewed one type of forecast through the survey. 

Following this set of introductions, participants went on to complete 12 decisions, which were presented in a random order. For each decision, they reported the binary decision to apply salt or not, and their confidence in this decision ranged from 50\% to 100\% \cite{tsai2008effects}. They also reported the likelihood that the temperature fell below a value that was less than the bottom limit of the distribution's range, which was used as an attention check. Each decision was also timed. Participants who spent less than 5 seconds on any given decision were also excluded from the analysis. 

After completing the 12 decisions, participants responded to a measure of overall trust in the type of forecast viewed throughout the survey. As in the work from Stokes et al., we used a multi-item measure for ``trust,'' consisting of usefulness, clarity, and accuracy \cite{elhamdadi2022we, pandey2023you, stokes2024delays, xiong2019examining}. Each item was rated on a scale from 0 (not clear/accurate/useful at all) to 10 (extremely clear/accurate/useful). The midpoint of the scale was a rating of 5 (moderately clear/accurate/useful).
In the analysis of trust, these ratings were averaged to provide a single value.

Participants also provided qualitative elaboration on their experience using the forecast, including aspects they liked and disliked about the information provided. For participants who were assigned to speech-forward forecasts, they answered an additional question asking if they took other actions while listening to the speech forecast. If they indicated that they were not sure if they did, they were provided examples (e.g., write anything down, draw anything, reflect aloud, etc.). If a participant responded that they did take additional actions, they were asked to provide a description.

In the final part of the survey, participants reported relevant demographics, including age range, education level, and gender. 
They also ranked the three relevant modes of information (visual, text, and speech) in order of preference. 
The final questions in this section were about participant experience with snow and ice, including how many years they had lived in an area where snow and ice were common during at least part of the year, how often they encountered icy road conditions during an average year, and whether they had ever been responsible for salting or de-icing roads or walkways.

\subsection{Hypotheses}

The hypotheses tested in this study sought to replicate the results from Stokes et al. by comparing speech, text, and visualization representations of uncertainty, as well as expand the results to consider variations of speech attributes. Hypotheses presented here are in line with the findings from Stokes et al. and prior work on acoustic variations and voice gender \cite{AhnKimSung2022, MooshammerEtzrodt2022, TolmeijerEtAl2021}.

This study explored combinations of visuals with other modes of information, particularly speech. These investigations are in part confirmatory and part exploratory, as we seek to replicate earlier findings on the matter while incorporating new variants to further explore the area of multimodal data uncertainty.
When making broad comparisons across modes, we condensed the speech-forward stimuli. We examined four attributes of decision-making: \textit{crossover temperature}, \textit{rationality} of decision (conservative, risky, or rational), decision \textit{confidence}, and overall \textit{trust} in the forecast. We examine three attributes of speech: gender, pitch (average pitch, normalized by gender), and speaking rate (duration of the first sentence and length of the first pause). Although the SSML adjustments described in \cref{sec:stimuli} all use the same delay duration, this attribute still varies slightly by voice.

\pheading{Crossover temperature.} Crossover temperature is the turning point at which participants were equally likely to salt or not salt. \textit{Optimal} crossover temperatures were calculated by determining the point at which a participant should start salting based on the cost/penalty framework of the decision. For this study, optimal crossover temperatures ranged from 32.3 to 32.5ºF. Our analysis focuses on how far participants tended to be, on average, from those crossover temperatures. To calculate this, we follow methods from prior work on decision-making with uncertainty \cite{padilla2021uncertain, stokes2024delays}.

\textbf{H1a:} Based on previous work, all three modes seem equally effective in communicating basic information about the data and eliciting attentive decisions from participants.  Thus, 
crossover temperatures will be similar between traditional visualization, text-forward, and speech-forward forecasts.

\textbf{H1b-d:} Men's voices, voices with a lower pitch, and voices with a faster speech rate tend to be perceived as more authoritative \cite{TolmeijerEtAl2021, MooshammerEtzrodt2022} and competent \cite{AhnKimSung2022}. Since there is little insight into the impact of gender and acoustic variations on decision quality, we expect that participants will make better decisions when they feel that they are getting good or reliable information from the speaker. 
Thus, crossover temperatures will be closer to optimal for men's voices than for women's voices \textbf{(b)}, for voices with a lower pitch than for those with a higher pitch \textbf{(c)}, and for voices with a faster rate of speaking than for those with a slower rate \textbf{(d)}. 

\pheading{Decision rationality.} Decisions for this framework could be categorized as conservative (i.e., applying salt even though the likelihood of freezing is low), risky (i.e., not applying salt even though the likelihood of freezing is high), or rational (i.e., applying salt appropriately). 
Stokes et al. \cite{stokes2024delays} found that speech forecasts tended to lead to more frequent risky decisions, possibly because speech feels like a more casual data representation or because it offer fewer specific likelihood details than visual representations.  
\textbf{H2a:} 
Decisions will be riskier for speech-forward forecasts than for traditional visualization forecasts.
\textbf{H2b-d:} Participants will make better decisions when they think the speaker is more reliable, and previous work suggests that perceived reliability is higher for men's voices and for voices with lower pitch and faster speech rate. 
Decisions will be more frequently rational for men's voices than for women's voices \textbf{(b)}, for voices with a lower pitch than for those with a higher pitch \textbf{(c)}, and for voices with a faster rate of speaking than for those with a slower rate \textbf{(d)}. 

\pheading{Decision confidence.} 
Stokes et al. found lower confidence for text forecasts in comparison to visualization and speech forecasts, possibly because readers are more confident that they can identify relevant information to inform their decision. 
\textbf{H3a}: 
Decision confidence will be lower for text-forward forecasts than for traditional visualization or speech-forward forecasts. 
\textbf{H3b-d:} Participants will be more confident in their decisions when they think the speaker is more reliable, and previous work suggests that perceived reliability is higher for men's voices and for voices with lower pitch and faster speech rate. 
Decision confidence will be higher for men's voices than for women's voices \textbf{(b)}, for voices with a lower pitch than for those with a higher pitch \textbf{(c)}, and for voices with a faster rate of speaking than for those with a slower rate \textbf{(d)}. 

\pheading{Trust.} 
\textbf{H4a}: Speech stimuli produce more of a social connection than visualizations or text, which can increase trust. 
Trust in forecasts will be higher for speech-forward stimuli than for text-forward stimuli or traditional visualizations.

Prior work indicates mixed findings regarding gender and trust \cite{GoodmanMayhorn2023, LeeNassBrave2000}. While women's voices tend to be perceived more positively in social characteristics such as warmth and sincerity \cite{AhnKimSung2022, MooshammerEtzrodt2022, TolmeijerEtAl2021}, men's voices tend to be perceived as having more authority \cite{AhnKimSung2022, TolmeijerEtAl2021}. The measure of trust is made up of perceived clarity, accuracy, and usefulness, and it is likely that these three attributes correspond more to the perception of reliability or authority than a social dynamic of trust.

\textbf{H4b-d:} Participants will have more faith in a speaker they think the speaker is more reliable, and previous work suggests that perceived reliability is higher for men's voices and for voices with lower pitch and faster speech rate. 
Trust will be higher for men's voices than for women's voices \textbf{(b)}, for voices with a lower pitch than for those with a higher pitch \textbf{(c)}, and for voices with a faster rate of speaking than for those with a slower rate \textbf{(d)}.

\section{Results}
\label{sec:results}

Participants tended to be successful at the task, with an average remaining budget of \$1,684 (average bonus of \$0.08). 
Time spent making the decisions was longer for speech-forward forecasts (mean = 51 seconds) than for text-forward (36s) or traditional visualization (29s) forecasts. This difference can be accounted for at least in part by the duration of the speech forecast, which averaged around 30 seconds. Participants typically only played the speech forecast one (69\%) or two (22\%) times for each trial. 

Participants ($n = 111$) generally found the forecasts easy to understand with appropriate detail. 
Some participants ($n = 27$) disliked the lack of the x-axis on the visual density mark provided, and others did not enjoy the sound of the specific voice ($n = 23$). These comments were relatively evenly spread across the different voices. 
A subset of participants ($n = 14$) took additional actions while listening to the speech forecast, with 12 participants writing down the numbers mentioned.

We further analyzed participant responses in terms of decision quality, confidence, and overall trust in the forecast \cite{rComputing}. When comparing models, we used ANOVA testing for model selection. Optimal models and significant findings (if present) are discussed. We used a $0.05$ cutoff for significance. When examining average pitch, we normalized this feature with respect to voice gender. 
Model tables can be found in supplemental materials \cite{stargazer2022package}. 

The findings tend to replicate those from Stokes et al. \cite{stokes2024delays}, with one major discrepancy. We replicated the findings that crossover temperature did not vary based on the mode of information, but speech led to more frequent risky decisions. We also replicated that decision confidence was lower for text-forward forecasts than for speech-forward forecasts. \textit{However, we did not find that speech-forward forecasts lead to higher trust.} This may be due to the bimodal representation compared to the unimodal representation tested by Stokes et al.

Beyond replication, we also examined several new hypotheses related to different elements of the speech forecasts themselves. The quality or rationality of the decisions did not vary between different acoustic features. Normalized pitch did have a small effect on decision confidence, but in the opposite direction as predicted. An increase in pitch led to an increase in decision confidence. Speech features had no effect on trust ratings.

\subsection{Decision Quality} 

\subsubsection{Crossover Temperature}

Hypotheses about crossover temperature were evaluated with logistic mixed effects models predicting the binary salting decision: ``Do not salt'' (0) or ``Salt'' (1). We compared models using ANOVA testing for model selection. Optimal models and significant findings (if present) are discussed. We used a $0.05$ cutoff for significance. 

\textbf{H1a:} Crossover temperatures will be similar between traditional visualization, text-forward, and speech-forward forecasts.

We found support for \textbf{H1a}. The optimal model for the salting decision was the baseline model, which only included the distance between the distribution mean and the optimal crossover temperature for the given distribution (crossover distance). This was compared to the optimal model from Stokes et al. \cite{stokes2024delays} as well as a forecast-specific model to compare specific voices to each other. We did not find that including any variables about participants' experiences with snow and ice improved model prediction.

Although differences between modes of information were not statistically significant, observed crossover temperatures for traditional visualization forecasts were further from optimal crossovers (0.38ºF) than for speech-forward forecasts (0.10ºF). Text-forward forecasts were in the middle (0.17ºF).

\textbf{H1b-d:} Crossover temperatures will be closer to optimal for men's voices than for women's voices \textbf{(b)}, for voices with a lower pitch than for those with a higher pitch \textbf{(c)}, and for voices with a faster rate of speaking than for those with a slower rate \textbf{(d)}. 

We did not find support for \textbf{H2b-d}. The optimal model for speech-specific salting decisions was also the baseline model. This model was compared to a series of models with increasing complexity, with the most complex model including an interaction between crossover distance and gender of the voice, the interaction between voice gender and participant gender \cite{LeeNassBrave2000}, and the three acoustic variables of interest (mean pitch, duration of the first sentence, and duration of the first pause). None of these variables improved the prediction of salting decisions.

\subsubsection{Decision Rationality}

\begin{figure}[ht]
    \centering
    \includegraphics[width = 0.95\linewidth]{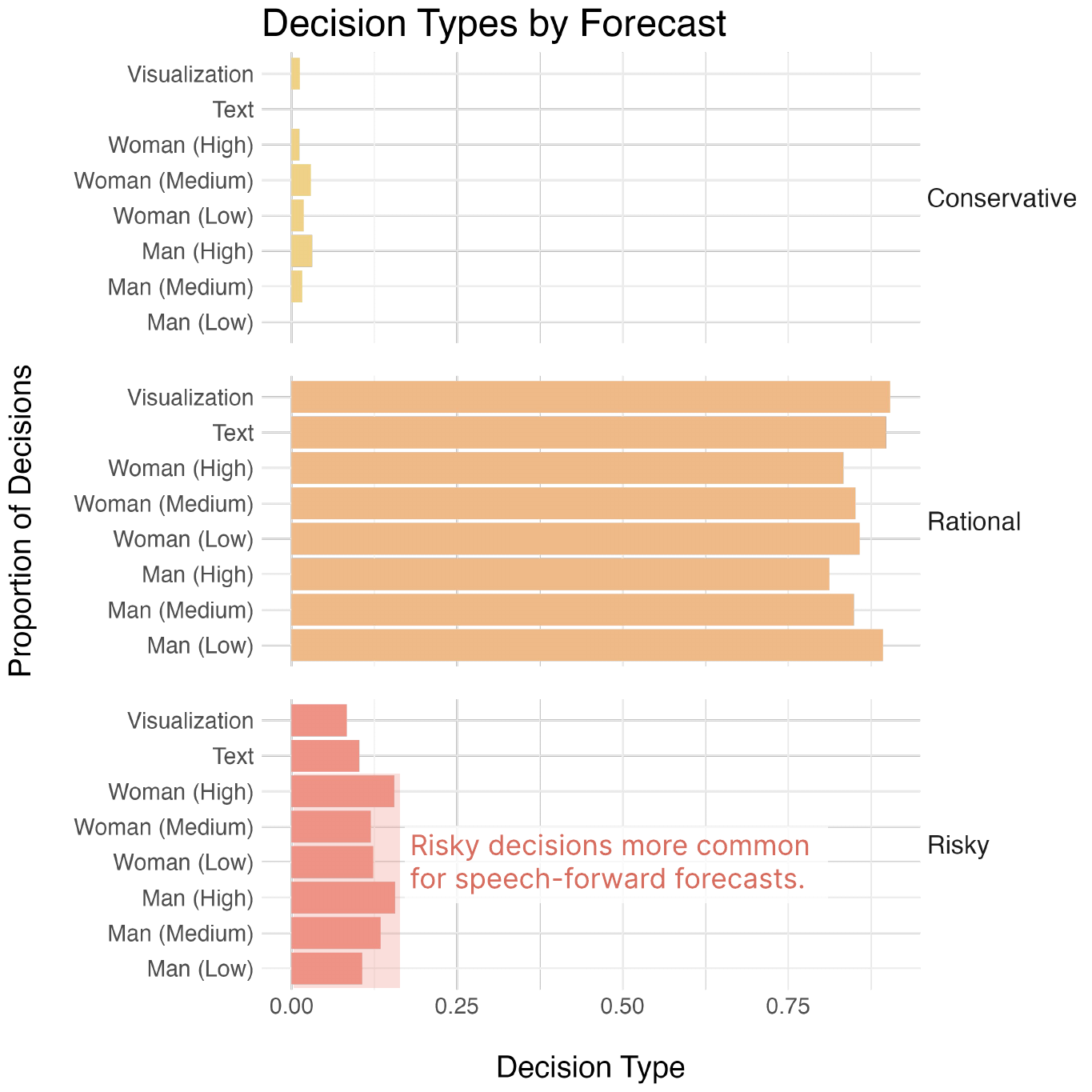}
    \caption{Proportion of decision types for each forecast. Overall, decisions were mostly rational. Speech-forward was the least rational representation, with a greater proportion of risky decisions. This observation was true across different voices as well.}  
    \label{fig:workshop_rationality}
\end{figure}

We evaluated decision rationality using $\chi^2$ tests when examining categorical variables such as mode and fitting a linear model and computing the ANOVA table when examining continuous variables. 
Exploratory testing indicated that there was no consistent effect of participant experiences with snow and ice on decision making rationality.
The frequency of decision types can be seen in \cref{fig:workshop_rationality}.

\textbf{H2a:} Decisions will be riskier for speech-forward forecasts than for traditional visualization forecasts.

We did find support for \textbf{H2a} ($\chi^2 = 30.1$, $p = 0.003$). 
We corroborate one of the main findings from Stokes et al. \cite{stokes2024delays}: speech forecasts led to more frequent risky decisions than visualization forecasts. Speech-forward forecasts had fewer rational decisions than expected ($standard residual = -3.5$) and more frequent risky ($SR = 2.9$) and conservative decisions ($SR = 2.1$) relative to expected. Text-forward forecasts had fewer conservative decisions than expected ($SR = -2.5$) and more frequent rational decisions ($SR = 2.0$). Traditional visualizations also had more frequent rational decisions than expected ($SR = 2.5$) and less frequent risky decisions ($SR = -2.5$). 

These values support \textbf{H2a} and the prior findings. Exploratory testing across the different speech-forward variants indicates that there was not a significant difference between decision-making across different voices ($\chi^2 = 17.2$, $p = 0.069$).

\textbf{H2b-d:} Decisions will be more frequently rational for men's voices than for women's voices \textbf{(b)}, for voices with a lower pitch than for those with a higher pitch \textbf{(c)}, and for voices with a faster rate of speaking than for those with a slower rate \textbf{(d)}.

We did not find support for \textbf{H2b-d}. 
There was no difference in decision rationality based on voice gender ($\chi^2 = 0.34$, $p = 0.844$). There was also no effect of pitch (Bonferroni-adjusted $p = 0.591$), the duration of the first sentence (Bonferroni-adjusted $p = 1.00$), nor the duration of the first pause (Bonferroni-adjusted $p = 0.641$)

\subsection{Confidence}
\begin{figure}[ht]
    \centering
    \includegraphics[width = \linewidth]{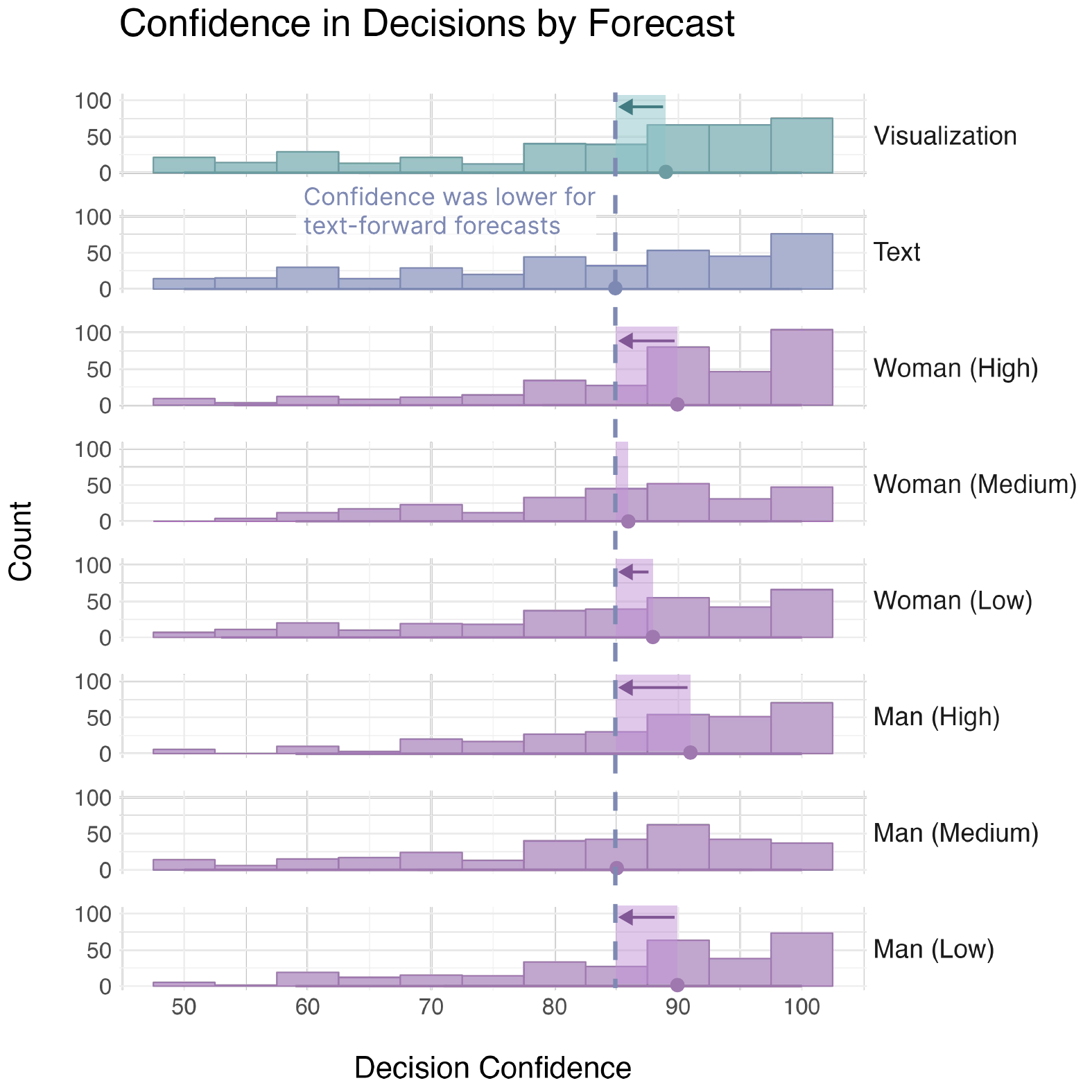}
    \caption{Confidence ratings ranged from 50 to 100. Overall, confidence was lower for text-forward ($mean = 82.4$) than for speech-forward forecasts ($mean = 85.5$).
    }
    \label{fig:workshop_confidence}
\end{figure}

Hypotheses about decision confidence were evaluated with linear mixed-effects models predicting the confidence rating, which ranged from 50 to 100. Distributions of confidence ratings for different forecast variants can be seen in \cref{fig:workshop_confidence}.

\textbf{H3a}: Decision confidence will be lower for text-forward forecasts than for traditional visualization or speech-forward forecasts. 

We found partial support for \textbf{H3a}. The optimal model ($p = 0.040$) included a random effect for participant and fixed effects of the difference between mean temperature and the optimal crossover, decision rationality, and mode of information.
Speech-forward forecasts led to higher confidence ratings compared to text-forward forecasts ($p = 0.031$), but there was no significant difference between text-forward forecasts and traditional visualization ($p = 0.673$). 
This effect was averaged over the different voices, and the model, which included forecast variant-specific fixed effects, did not improve prediction.

\textbf{H3b:} Decision confidence will be higher for men's voices than for women's voices.

We did not find support for \textbf{H3b}. 
The optimal model ($p < 0.001$) for speech-specific hypotheses included a random effect for participant and fixed effects of the crossover distance, decision rationality, and acoustic features. However, it did not include voice gender, thus providing no support for \textbf{H3b}. 

\textbf{H3c:} Decision confidence will be higher for voices with a lower pitch than for those with a higher pitch. 

We did not find support for \textbf{H3c}. Normalized pitch did have a significant effect in our model ($p = 0.021$) but in the opposite direction. An increase of one standard deviation in normalized pitch would result in a 2.01-point increase [1.15, 2.87] in confidence ratings. This effect is about 4\% of the confidence scale.

\textbf{H3d:} Decision confidence will be higher for voices with a faster rate of speaking than for those with a slower rate. 

We did not find support for \textbf{H3d}. Neither measure of the rate of speaking had an effect on decision confidence.

\subsection{Trust}

Hypotheses about trust in forecasts were evaluated with linear models predicting the average trust rating, which ranged from 0 to 10. Distributions of trust ratings for different forecast variants can be seen in \cref{fig:workshop_confidence}.

\begin{figure}[ht]
    \centering
    \includegraphics[width = \linewidth]{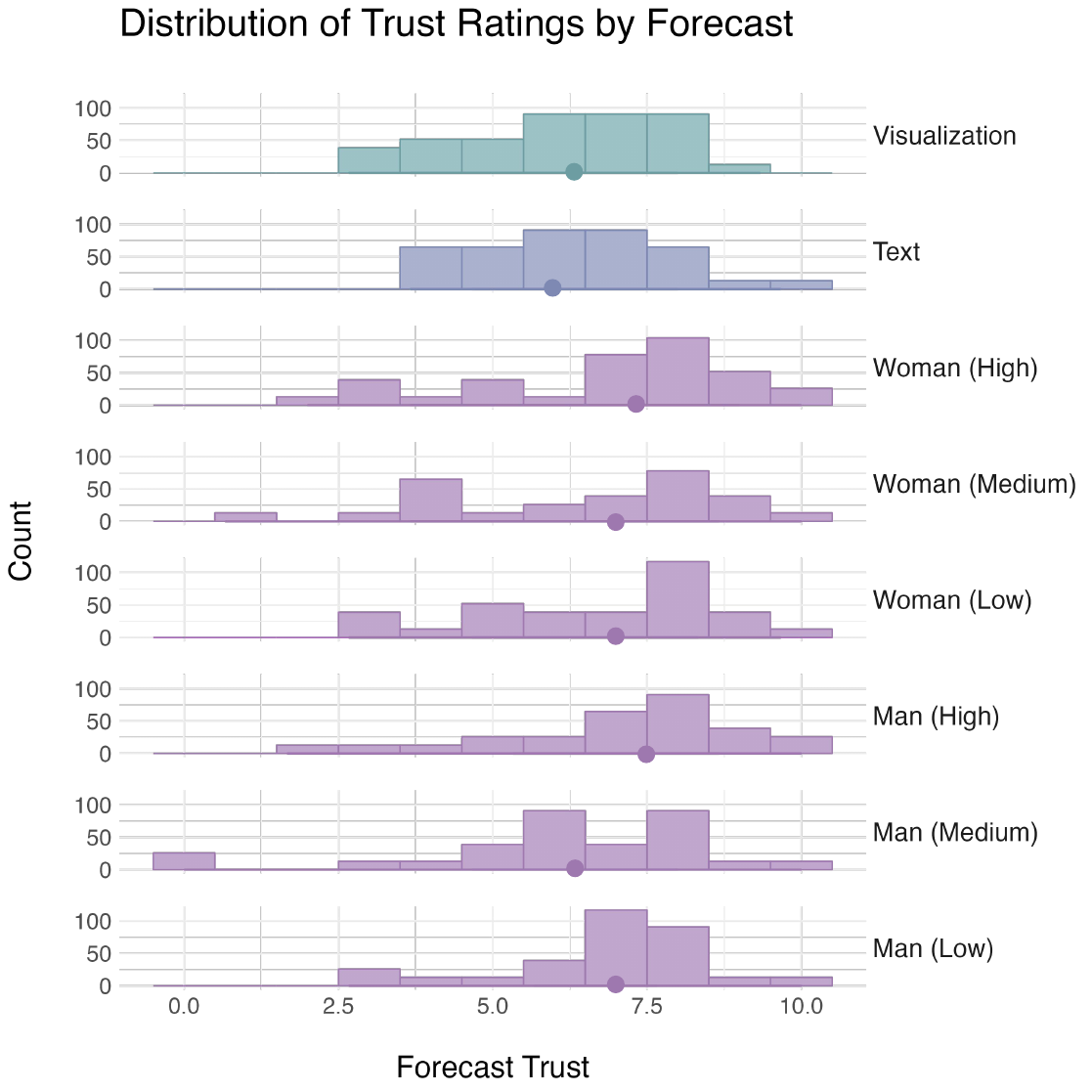}
    \caption{Average trust ratings. There were no significant differences between modes, but ratings were higher overall for speech-forward forecasts ($mean = 6.7$) than for text-forward ($mean = 6.3$) or traditional visualization ($mean = 6.1$).
    }
    \label{fig:workshop_trust}
\end{figure}

\textbf{H4a}: Trust in forecasts will be higher for speech-forward stimuli than for text-forward stimuli or traditional visualizations. 

We did not find support for \textbf{H4a}. Contrary to the findings from Stokes et al., the optimal model was the baseline model, which only included a fixed effect of the average rationality of the participant. Neither mode nor forecast-specific effects improved prediction significantly. It is possible that the bimodal presentation (compared to the original unimodal representation) affected perceptions of trust. We explore this further in \cref{sec:discuss}.

\textbf{H4b-d:} Trust will be higher for men's voices than for women's voices \textbf{(b)}, for voices with a lower pitch than for those with a higher pitch \textbf{(c)}, and for voices with a faster rate of speaking than for those with a slower rate \textbf{(d)}. 

We did not find support for \textbf{H4b-d}. Again, the optimal model for the speech-specific hypotheses was the baseline model. We did not see an improvement in prediction by including gender or acoustic features. It is possible that the measure of trust used was evaluating elements not affected strongly by specific features of speech. 

\section{Discussion}
\label{sec:discuss}

Overall, the results of this study support prior insights \cite{stokes2024delays} into the impact of different modes of information on decision-making under uncertainty. Although the number of voices tested (six) was likely too small to make broad generalizations about acoustic features, testing additional speech variants allows us to follow up on the open questions left by prior work regarding the generalizability of findings and the impact of voice gender on decision-making. 

\textbf{Speech-forward forecasts led to more frequent risky decisions than text-forward or visualization-only forecasts}. 
This reduced rationality may be due to a number of factors, including the transient nature of speech and the effort required to hold information in working memory; only a small subset of participants (8\%) wrote down information from the forecast. The increase in risky decisions could also be due to aspects of the forecast that appear emphasized in speech but not in text or visual representations. 
We did not observe significant differences between different voices or acoustic features, nor for different levels of snow experiences, although there was variation between speech and the other modes, as shown in \cref{fig:workshop_rationality}. 
Variations observed between voices may be natural variations in task performance between participant groups or may be influenced by aspects of speech that were not tested here.

\textbf{Confidence in decisions when using text-forward forecasts is lower compared to speech-forward forecasts.} This finding is interesting when considering that text-forward forecasts lead to more frequent rational decisions than speech-forward forecasts and that they offer the same information in the same words.  
This difference could be interpreted as either a \textit{deflation} of confidence when using text or an \textit{inflation} of confidence when using speech. 
The inclusion of the visual mark did not increase confidence for text condition in comparison to unimodal representations \cite{stokes2024delays}. 

When comparing speech-forward forecasts, \textbf{increased normalized pitch led to increased confidence in the decision.} Due to a limited number of voices tested, it is difficult to make a generalized interpretation of this result. While previous work has found that voices with lower pitch are perceived as more authoritative and competent \cite{AhnKimSung2022, MooshammerEtzrodt2022, TolmeijerEtAl2021}, this task asked participants to evaluate their confidence in their own decisions. In this context, the socially positive perception of higher-pitched voices \cite{TolmeijerEtAl2021} may increase participants' confidence.  

It is also possible that listeners were less engaged with the lower-pitched voices because those voices had some characteristics of creaky phonation, which is perceived negatively \cite{AndersonEtAl2014}. Another potential influence is the realization of /t/ between vowels.  The voices with the highest confidence ratings produced this consonant with a full closure and aspirated release ([t\textsuperscript{h}]), while the speakers with the lowest confidence ratings produced it as a flap ([\textipa{R}]).
Other differences between the voices might also be responsible for the results, given that pitch was not the only characteristic that differed.  For example, \textsc{Man Medium} and \textsc{Woman Medium} have lower formants (resonant frequencies) than the other voices.

\textbf{Ratings of trust did not differ between the three modes of information.} This result is in conflict with prior findings that speech forecasts led to higher ratings of trust. It is possible that the bimodal representations tested in this experiment affected the clarity, accuracy, and usefulness of the forecast. 11\% of participants commented that they disliked the visual mark or found it confusing to interpret. The average trust rating for the bimodal speech-forward forecast ($6.7$) was almost a full point lower than the rating for the unimodal speech forecast (7.6) tested in prior work \cite{stokes2024delays}, indicating there may have been an impact of introducing visual features.

Additionally, \textbf{gender and acoustic features did not have an effect on trust ratings.}
Although previous work has found effects of voice gender and specific acoustic characteristics on explicit evaluations of the voice \cite{AhnKimSung2022, GoodmanMayhorn2023, LeeNassBrave2000, MooshammerEtzrodt2022, OleszkiewiczEtAl2017, TolmeijerEtAl2021} as well as engagement with advertisements \cite{AhnKimSung2022, EfthymiouEtAl2024} and decisions about described scenarios \cite{GoodmanMayhorn2023, LeeNassBrave2000}, those biases might be outweighed in contexts where behavioral information about the speaker's reliability is available \cite{knight:2021}. 
In our study, the presence of the density mark might have contributed to establishing all voices as accurately describing the data. 
Gender biases might be more apparent when listeners make decisions that are more subjective or not accompanied by supporting visualizations of the data.


\begin{figure}[!ht]
    \centering
    \includegraphics[width = 0.9\linewidth]{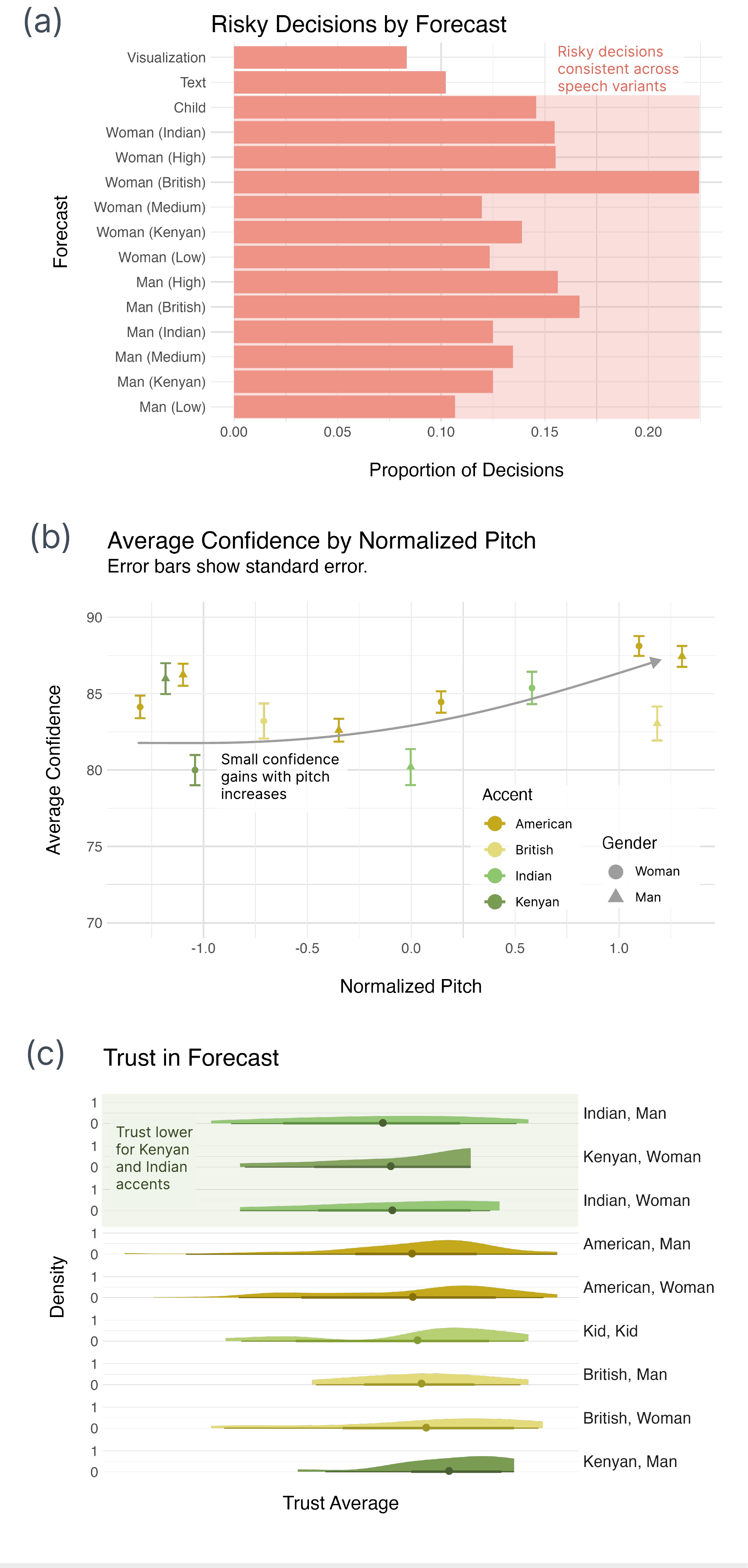}
    \caption{Findings from an exploratory investigation of accents. (a) Decision rationality between different voices tested. Risky decisions were consistently more common for speech-forward forecasts but did not vary by accent. (b) Confidence ratings and normalized pitch. Accent variants continued the minor trend observed in the experiment; there were small gains in confidence for increases in pitch. (c) Trust in forecast by accent and gender. Indian and Kenyan accents tended to have the lowest average trust.
    }  
    \label{fig:exploratory}
\end{figure}

\subsection{Exploratory Study of Additional Speech Variants}
Overall, findings were typically consistent across gender and acoustic features. However, there are other aspects of voices that were not examined in this study, including accent and age. 
These characteristics might impact comprehension, trust, and decision-making due to varying degrees of familiarity and associated biases. Additionally, as many virtual assistant technologies (e.g., Alexa, Siri) can use different accents, understanding their effects is increasingly relevant.
We conducted a smaller, exploratory study to compare additional speech-forward forecasts, focusing primarily on comparisons of different accents.  

In this exploratory study, we tested seven additional speech variants: a child's voice, a British accent, a Kenyan accent, and an Indian accent. For each accent, we tested corresponding man and woman voices. We could not test multiple variants of each combination of gender and accent due to limitations in the Microsoft Azure options.
Speech stimuli were created using the same method described in \cref{sec:speech_links}. The only change made to the study design was the inclusion of measures of accent familiarity, including the identification of the accent, report of personal experience, and a familiarity rating \cite{huang2016cross}. 
We recruited 105 participants from Prolific \cite{palan2018prolific}, with 90 responses remaining after exclusions based on attention checks. 
Since this investigation was exploratory, we did not complete any significance testing or examine any specific hypotheses. Instead, we examined general trends and compared the new variants to the original study conditions, all six of which had American accents. 
Interesting findings from this exploration are shown in \cref{fig:exploratory}.
Further details can be found in supplementary materials. 

Overall, participants were most familiar with Indian and British accents. 
79\% of participants could identify Indian accents, and 38\% had personal experience with someone who spoke with that accent. Similarly, 75\% of participants could identify British accents, and 38\% had personal experience. Participants tended to be less familiar with the Kenyan accent, with only 29\% of participants able to identify it and 13\% with personal experience.

The higher proportion of risky decisions for speech-forward forecasts than for other modes was consistent across different accents, including the child's voice.  There was some variation in decision riskiness across individual voices, but it did not fall into clear patterns.
Rationality between the different conditions tended to be similar.
The connection between increased confidence ratings and increased normalized pitch was consistent when including these variants as well, but we also did not observe a strong effect of accent on its own. 

Each component of the trust scores varied substantially across voices. Both Indian voices and the Kenyan woman voice
scored lower on clarity, accuracy, and usefulness compared to American and British accents. However, the Kenyan man voice did not.  One potential reason for this observation is that this voice had the fastest speech rate of any variant tested; previous work shows that faster speech is perceived as more credible and more persuasive \cite{MehrabianWilliams1969, MillerEtAl1976}. While the Indian voices had slower speaking rates, they were comparable to several other voices (\textsc{Woman Low}, \textsc{Woman British}, and \textsc{Kid}); speech rate is not the only factor influencing these scores. The effects of speech rate might be partially obscured by other variations across the voices, given the limited set of variants.

This exploration helps to further investigate how different speech variants affect aspects of decision-making under uncertainty, illuminating possible effects on decision confidence and trust.
The findings of the main experiment also support prior work, expanding our understanding of decision-making with different modes of decision to more generalizable findings. 

\section{Limitations and Future Work}
\label{sec:limits_fw}

The study was limited by the availability of synthetic voice options. Even within the American English category tested in our main experiment, the range of acoustic characteristics was narrow;
the limited variation across voices imposed constraints on testing effects of acoustic variables. Furthermore, control over independent voice parameters was limited, which meant that discerning the specific impact of various acoustic features on perception was speculative. 

Additional complexities may have arisen from comprehensibility issues associated with unfamiliar or foreign accents, potentially introducing a bias against out-group members~\cite{FoucartEtAl2020}. 
The stimuli also comprised binary gender voices that may not fully represent the diversity of voice perceptions. While there is some previous research on the perception of nonbinary or gender-ambiguous voices \cite{TolmeijerEtAl2021,MooshammerEtzrodt2022}, future work needs to explore the potential effects of gender-ambiguous voices in communicating data uncertainty. 

Our decision to remove axes from the visualization in the “text-forward” and “speech-forward” conditions was unconventional in standard visualization practices and so may pose a limitation for these findings. This choice allowed us to better isolate the role of text and speech but may limit the generalizability of our findings. Future work in multimodal communication should examine a variety of visual representations, including the full visualization.

Further, the extent to which the observed effects translate to tasks of differing personal importance or interest is unknown. For example, the effects might differ in contexts like driving instructions, advertising, or opinion-based tasks such as political commentary, where the personal stakes or engagement levels may alter how speech is perceived and acted upon. The study findings also identify future research directions as well as implications for practical domains in exploring the implications of voice in communicating uncertainty across various domains and user needs.

\pheading{Accessibility and personalization.} For individuals with visual impairments, rapid communication of textual information trades off with decreased comprehensibility of fast speech produced by text-to-speech systems \cite{StentSyrdalMishra2011}. 
Combining voice-driven systems with visual aids has the potential to significantly aid in visualizing data uncertainty. For instance, a voice assistant could explain complex graphical data, such as statistical uncertainty or probability distributions, while the visual aid provides a graphical representation. This dual-mode delivery can cater to users with different sensory preferences or disabilities, ensuring that the information is accessible to all. The study did not account for individual user preferences in choosing voice options, which could significantly affect the outcomes. If users had the freedom to select voices, they might opt for those that align more closely with their personal preferences or the specific context of the task, potentially leading to different decision-making outcomes than those observed under controlled experimental conditions.

\pheading{Cross-cultural communication.} Different cultures may have distinct preferences and interpretations regarding voice characteristics such as tone, pitch, speed, and accent; what is considered a trustworthy voice in one culture may be perceived differently in another~\cite{dehghani:2014}. Understanding these variations is crucial to help guide the customization of voice attributes to align with cultural expectations, reducing uncertainty and improving clarity in communication. The perception of authority and trustworthiness through voice could affect how information, particularly data that involves uncertainty, is received by an audience \cite{TolmeijerEtAl2021}.

\pheading{Advertising.} In the context of communicating data uncertainty, advertising could play an important role by using strategic voice characteristics to enhance trust, clarify ambiguities, and simplify complex information~\cite{Dubey2018}. Effective advertising can demystify uncertainties associated with products like financial services, using a trustworthy voice to reassure consumers and manage expectations. Future research could explore how different voice characteristics with corresponding data and visualizations influence consumer behavior in advertising (cf. \cite{AhnKimSung2022, EfthymiouEtAl2024} for effects of voice characteristics). By manipulating variables such as pitch, speed, and accent, market studies can determine which acoustic features most effectively persuade different demographic groups. 

\pheading{News and journalism.} The visual and vocal attributes of news delivery play a significant role in how information is perceived and trusted by the public. Different types of news might benefit from specific voice attributes to match the content's nature and urgency. When reporting on stories with inherent uncertainties, such as weather predictions, economic projections, or evolving health crises, the voice delivering the news may significantly affect how the information is received. 

\pheading{Long-term impact of voice on uncertainty communication.} To understand the long-term effects of voice characteristics on user behavior and perception, longitudinal studies could be conducted. These studies would provide insights into how consistent exposure to certain voice types might influence user trust, satisfaction, and loyalty over time. Longitudinal studies can also further investigate how users adapt to and learn from voice-guided systems over time. This exploration could include how users' comprehension of and responses to communicate uncertainties improve as they become more familiar with a specific voice's acoustical characteristics.

\section{Conclusion}
\label{sec:conclusion}

This study examined how speaker characteristics and acoustic variables impact decision-making in the context of uncertainty communication. The findings suggest that the modality of information—whether through speech, text, or a combination—significantly influences how it is used and perceived. While voices with the same accent showed minimal variation, certain acoustic features, such as pitch, may still affect decision confidence. Speech-forward conditions led to more frequent risky decisions, emphasizing the importance of carefully designing speech attributes in communication tools. Text-forward stimuli consistently resulted in lower confidence compared to speech, indicating a need for improved text utilization in multimodal strategies. Although the study did not find significant differences in trust based on voice gender or acoustic characteristics among American voices, exploratory analysis suggests that accents could affect listener perception. These insights identify research directions for further exploring the nuances around \textit{expressing} uncertainty across various multimodal contexts and applications.






\bibliographystyle{abbrv-doi-narrow}

\bibliography{bibliography}

\begin{thebibliography}{10}
\renewcommand*{\sfdefault}{PTSansNarrow-TLF}

\bibitem{AhnKimSung2022}
J.~Ahn, J.~Kim, and Y.~Sung.
\newblock {The Effect of Gender Stereotypes on Artificial Intelligence Recommendations}.
\newblock {\em Journal of Business Research}, 141:50--59, 2022. doi: \textsf{%
10\hspace{.1pt}\discretionary{.}{%
}{.}\hspace{.4pt}1016\discretionary{/}{%
}{/}j\hspace{.1pt}\discretionary{.}{%
}{.}\hspace{.4pt}jbusres\hspace{.1pt}\discretionary{.}{%
}{.}\hspace{.4pt}2021\hspace{.1pt}\discretionary{.}{%
}{.}\hspace{.4pt}12\hspace{.1pt}\discretionary{.}{%
}{.}\hspace{.4pt}007}


\bibitem{AndersonEtAl2014}
R.~C. Anderson, C.~A. Klofstad, W.~J. Mayew, and M.~Venkatachalam.
\newblock {Vocal Fry May Undermine the Success of Young Women in the Labor Market}.
\newblock {\em PloS One}, 9(5):e97506, 2014. doi: \textsf{%
10\hspace{.1pt}\discretionary{.}{%
}{.}\hspace{.4pt}1371\discretionary{/}{%
}{/}journal\hspace{.1pt}\discretionary{.}{%
}{.}\hspace{.4pt}pone\hspace{.1pt}\discretionary{.}{%
}{.}\hspace{.4pt}0097506}


\bibitem{bancilhon2023combining}
M.~Bancilhon, A.~Wright, S.~Ha, R.~J. Crouser, and A.~Ottley.
\newblock {Why Combining Text and Visualization Could Improve Bayesian Reasoning: A Cognitive Load Perspective}.
\newblock In {\em CHI Conference on Human Factors in Computing Systems}, pp. 1--15, 2023.

\bibitem{BrennanWilliams1995}
S.~E. Brennan and M.~Williams.
\newblock The {F}eeling of {A}nother's {K}nowing: {P}rosody and {F}illed {P}auses as {C}ues to {L}isteners {A}bout the {M}etacognitive {S}tates of {S}peakers.
\newblock {\em Journal of Memory and Language}, 34:383--398, 1995. doi: \textsf{%
10\hspace{.1pt}\discretionary{.}{%
}{.}\hspace{.4pt}1006\discretionary{/}{%
}{/}jmla\hspace{.1pt}\discretionary{.}{%
}{.}\hspace{.4pt}1995\hspace{.1pt}\discretionary{.}{%
}{.}\hspace{.4pt}1017}


\bibitem{BrownGilesThakerar1985}
B.~L. Brown, H.~Giles, and J.~N. Thakerar.
\newblock {Speaker Evaluations as a Function of Speech Rate, Accent and Context.}
\newblock {\em Language \& Communication}, 1985. doi: \textsf{%
10\hspace{.1pt}\discretionary{.}{%
}{.}\hspace{.4pt}1016\discretionary{/}{%
}{/}0271\discretionary{%
}{-}{-}5309\discretionary{%
}{(}{(}85\discretionary{)}{%
}{)}90011\discretionary{%
}{-}{-}4}


\bibitem{CapurroEtAl2021}
G.~Capurro, C.~G. Jardine, J.~Tustin, and M.~Driedger.
\newblock Communicating {S}cientific {U}ncertainty in a {R}apidly {E}volving {S}ituation: {A} {F}raming {A}nalysis of {C}anadian {C}overage in {E}arly {D}ays of {COVID-19}.
\newblock {\em BMC Public Health}, 21(1):2181, 2021. doi: \textsf{%
10\hspace{.1pt}\discretionary{.}{%
}{.}\hspace{.4pt}1186\discretionary{/}{%
}{/}s12889\discretionary{%
}{-}{-}021\discretionary{%
}{-}{-}12246\discretionary{%
}{-}{-}x}


\bibitem{correll2014error}
M.~Correll and M.~Gleicher.
\newblock {Error Bars Considered Harmful: Exploring Alternate Encodings for Mean and Error}.
\newblock {\em IEEE Transactions on Visualization and Computer Graphics}, 20(12):2142--2151, 2014.

\bibitem{dehghani:2014}
M.~Dehghani, P.~Khooshabeh, A.~Nazarian, and J.~Gratch.
\newblock {The Subtlety of Sound}.
\newblock {\em Journal of Language and Social Psychology}, 34:0261927X14551095, 05 2014. doi: \textsf{%
10\hspace{.1pt}\discretionary{.}{%
}{.}\hspace{.4pt}1177\discretionary{/}{%
}{/}0261927X14551095}


\bibitem{draper:1995}
D.~Draper.
\newblock Assessment and {P}ropagation of {M}odel {U}ncertainty.
\newblock {\em Journal of the Royal Statistical Society, Series B (Methodological)}, 57:45--97, 1995. doi: \textsf{%
10\hspace{.1pt}\discretionary{.}{%
}{.}\hspace{.4pt}2307\discretionary{/}{%
}{/}2346087}


\bibitem{Dubey2018}
M.~Dubey, J.~Farrell, and L.~Ang.
\newblock {How Accent and Pitch Affect Persuasiveness in Radio Advertising}.
\newblock In V.~Cauberghe, L.~Hudders, and M.~Eisend, eds., {\em Advances in Advertising Research IX: Power to Consumers}, pp. 117--130. Springer Fachmedien Wiesbaden, Wiesbaden, 2018. doi: \textsf{%
10\hspace{.1pt}\discretionary{.}{%
}{.}\hspace{.4pt}1007\discretionary{/}{%
}{/}978\discretionary{%
}{-}{-}3\discretionary{%
}{-}{-}658\discretionary{%
}{-}{-}22681\discretionary{%
}{-}{-}7\_9}


\bibitem{EfthymiouEtAl2024}
F.~Efthymiou, C.~Hildebrand, E.~de~Bellis, and W.~Hampton.
\newblock {The Power of {AI}-generated Voices: How Digital Vocal Tract Length Shapes Product Congruency and Ad Performance}.
\newblock {\em Journal of Interactive Marketing}, 59(2):117--134, 2024. doi: \textsf{%
10\hspace{.1pt}\discretionary{.}{%
}{.}\hspace{.4pt}1177\discretionary{/}{%
}{/}10949968231194905}


\bibitem{elhamdadi2022we}
H.~Elhamdadi, A.~Gaba, Y.-S. Kim, and C.~Xiong.
\newblock How {D}o {W}e {M}easure {T}rust in {V}isual {D}ata {C}ommunication?
\newblock In {\em 2022 IEEE Evaluation and Beyond-Methodological Approaches for Visualization (BELIV)}, pp. 85--92. IEEE, Piscataway, NJ, 2022. doi: \textsf{%
10\hspace{.1pt}\discretionary{.}{%
}{.}\hspace{.4pt}1109\discretionary{/}{%
}{/}BELIV57783\hspace{.1pt}\discretionary{.}{%
}{.}\hspace{.4pt}2022\hspace{.1pt}\discretionary{.}{%
}{.}\hspace{.4pt}00014}


\bibitem{faul2009statistical}
F.~Faul, E.~Erdfelder, A.~Buchner, and A.-G. Lang.
\newblock Statistical {P}ower {A}nalyses using {G}* {P}ower 3.1: {T}ests for {C}orrelation and {R}egression {A}nalyses.
\newblock {\em Behavior Research Methods}, 41(4):1149--1160, 2009. doi: \textsf{%
10\hspace{.1pt}\discretionary{.}{%
}{.}\hspace{.4pt}3758\discretionary{/}{%
}{/}BRM\hspace{.1pt}\discretionary{.}{%
}{.}\hspace{.4pt}41\hspace{.1pt}\discretionary{.}{%
}{.}\hspace{.4pt}4\hspace{.1pt}\discretionary{.}{%
}{.}\hspace{.4pt}1149}


\bibitem{faul2007g}
F.~Faul, E.~Erdfelder, A.-G. Lang, and A.~Buchner.
\newblock G* {P}ower 3: {A} {F}lexible {S}tatistical {P}ower {A}nalysis {P}rogram for the {S}ocial, {B}ehavioral, and {B}iomedical {S}ciences.
\newblock {\em Behavior Research Methods}, 39(2):175--191, 2007. doi: \textsf{%
10\hspace{.1pt}\discretionary{.}{%
}{.}\hspace{.4pt}3758\discretionary{/}{%
}{/}BF03193146}


\bibitem{FeldsteinDohmCrown2001}
S.~Feldstein, F.-A. Dohm, and C.~L. Crown.
\newblock {Gender and Speech Rate in the Perception of Competence and Social Attractiveness}.
\newblock {\em The Journal of Social Psychology}, 141(6):785--806, 2001. doi: \textsf{%
10\hspace{.1pt}\discretionary{.}{%
}{.}\hspace{.4pt}1080\discretionary{/}{%
}{/}00224540109600588}


\bibitem{fernandes2018uncertainty}
M.~Fernandes, L.~Walls, S.~Munson, J.~Hullman, and M.~Kay.
\newblock Uncertainty {D}isplays {U}sing {Q}uantile {D}otplots or {CDF}s {I}mprove {T}ransit {D}ecision-{M}aking.
\newblock In {\em Proceedings of the 2018 CHI Conference on Human Factors in Computing Systems}, pp. 1--12. Association for Computing Machinery, New York, NY, USA, 2018. doi: \textsf{%
10\hspace{.1pt}\discretionary{.}{%
}{.}\hspace{.4pt}1145\discretionary{/}{%
}{/}3173574\hspace{.1pt}\discretionary{.}{%
}{.}\hspace{.4pt}317371}


\bibitem{FoucartEtAl2020}
A.~Foucart, A.~Costa, L.~Mor{\'\i}s-Fern{\'a}ndez, and R.~J. Hartsuiker.
\newblock {Foreignness or Processing Fluency? On Understanding the Negative Bias Toward Foreign-accented Speakers}.
\newblock {\em Language Learning}, 70(4):974--1016, 2020. doi: \textsf{%
10\hspace{.1pt}\discretionary{.}{%
}{.}\hspace{.4pt}1111\discretionary{/}{%
}{/}lang\hspace{.1pt}\discretionary{.}{%
}{.}\hspace{.4pt}12413}


\bibitem{franconeri2021science}
S.~L. Franconeri, L.~M. Padilla, P.~Shah, J.~M. Zacks, and J.~Hullman.
\newblock {The Science of Visual Data Communication: What Works}.
\newblock {\em Psychological Science in the Public Interest}, 22(3):110--161, 2021. doi: \textsf{%
10\hspace{.1pt}\discretionary{.}{%
}{.}\hspace{.4pt}1177\discretionary{/}{%
}{/}15291006211051956}


\bibitem{FuertesEtAl2012}
J.~N. Fuertes, W.~H. Gottdiener, H.~Martin, T.~C. Gilbert, and H.~Giles.
\newblock {A Meta-analysis of the Effects of Speakers' Accents on Interpersonal Evaluations}.
\newblock {\em European Journal of Social Psychology}, 42(1):120--133, 2012. doi: \textsf{%
10\hspace{.1pt}\discretionary{.}{%
}{.}\hspace{.4pt}1002\discretionary{/}{%
}{/}ejsp\hspace{.1pt}\discretionary{.}{%
}{.}\hspace{.4pt}862}


\bibitem{FurnhamGunterGreen1990}
A.~Furnham, B.~Gunter, and A.~Green.
\newblock Remembering {S}cience: {T}he {R}ecall of {F}actual {I}nformation as a {F}unction of the {P}resentation {M}ode.
\newblock {\em Applied Cognitive Psychology}, 4(3):203--212, 1990. doi: \textsf{%
10\hspace{.1pt}\discretionary{.}{%
}{.}\hspace{.4pt}1002\discretionary{/}{%
}{/}acp\hspace{.1pt}\discretionary{.}{%
}{.}\hspace{.4pt}2350040305}


\bibitem{galesic2009using}
M.~Galesic, R.~Garcia-Retamero, and G.~Gigerenzer.
\newblock {Using Icon Arrays to Communicate Medical Risks: Overcoming Low Numeracy.}
\newblock {\em Health Psychology}, 28(2):210, 2009.

\bibitem{Giles1971}
H.~Giles.
\newblock {Patterns of Evaluation to {RP}, {South Welsh} and {Somerset} Accented Speech}.
\newblock {\em British Journal of Social and Clinical Psychology}, 10(3):280--281, 1971. doi: \textsf{%
10\hspace{.1pt}\discretionary{.}{%
}{.}\hspace{.4pt}1111\discretionary{/}{%
}{/}j\hspace{.1pt}\discretionary{.}{%
}{.}\hspace{.4pt}2044\discretionary{%
}{-}{-}8260\hspace{.1pt}\discretionary{.}{%
}{.}\hspace{.4pt}1971\hspace{.1pt}\discretionary{.}{%
}{.}\hspace{.4pt}tb00748\hspace{.1pt}\discretionary{.}{%
}{.}\hspace{.4pt}x}


\bibitem{GoodmanMayhorn2023}
K.~L. Goodman and C.~B. Mayhorn.
\newblock {It's Not What You Say But How You Say It: Examining the Influence of Perceived Voice Assistant Gender and Pitch on Trust and Reliance}.
\newblock {\em Applied Ergonomics}, 106:103864, 2023. doi: \textsf{%
10\hspace{.1pt}\discretionary{.}{%
}{.}\hspace{.4pt}1016\discretionary{/}{%
}{/}j\hspace{.1pt}\discretionary{.}{%
}{.}\hspace{.4pt}apergo\hspace{.1pt}\discretionary{.}{%
}{.}\hspace{.4pt}2022\hspace{.1pt}\discretionary{.}{%
}{.}\hspace{.4pt}103864}


\bibitem{stargazer2022package}
M.~Hlavac.
\newblock {\em {S}targazer: {W}ell-{F}ormatted {R}egression and {S}ummary {S}tatistics {T}ables}.
\newblock Social Policy Institute, Bratislava, Slovakia, 2022.
\newblock R package version 5.2.3.

\bibitem{HuPan2024}
Q.~Hu and Z.~Pan.
\newblock {Is Cute {AI} More Forgivable? The Impact of Informal Language Styles and Relationship Norms of Conversational Agents on Service Recovery}.
\newblock {\em Electronic Commerce Research and Applications}, 65:101398, 2024. doi: \textsf{%
10\hspace{.1pt}\discretionary{.}{%
}{.}\hspace{.4pt}1016\discretionary{/}{%
}{/}j\hspace{.1pt}\discretionary{.}{%
}{.}\hspace{.4pt}elerap\hspace{.1pt}\discretionary{.}{%
}{.}\hspace{.4pt}2024\hspace{.1pt}\discretionary{.}{%
}{.}\hspace{.4pt}101398}


\bibitem{huang2016cross}
B.~Huang, A.~Alegre, and A.~Eisenberg.
\newblock {A Cross-linguistic Investigation of the Effect of Raters' Accent Familiarity on Speaking Assessment}.
\newblock {\em Language Assessment Quarterly}, 13(1):25--41, 2016. doi: \textsf{%
10\hspace{.1pt}\discretionary{.}{%
}{.}\hspace{.4pt}1080\discretionary{/}{%
}{/}15434303\hspace{.1pt}\discretionary{.}{%
}{.}\hspace{.4pt}2015\hspace{.1pt}\discretionary{.}{%
}{.}\hspace{.4pt}1134540}


\bibitem{hullman2020why}
J.~Hullman.
\newblock {Why Authors Don't Visualize Uncertainty}.
\newblock {\em IEEE Transactions on Visualization and Computer Graphics}, 26(1):130--139, 2020. doi: \textsf{%
10\hspace{.1pt}\discretionary{.}{%
}{.}\hspace{.4pt}1109\discretionary{/}{%
}{/}TVCG\hspace{.1pt}\discretionary{.}{%
}{.}\hspace{.4pt}2019\hspace{.1pt}\discretionary{.}{%
}{.}\hspace{.4pt}2934287}


\bibitem{joslyn2012uncertainty}
S.~L. Joslyn and J.~E. LeClerc.
\newblock Uncertainty {F}orecasts {I}mprove {W}eather-{R}elated {D}ecisions and {A}ttenuate the {E}ffects of {F}orecast {E}rror.
\newblock {\em Journal of Experimental Psychology: Applied}, 18(1):126--140, 2012. doi: \textsf{%
10\hspace{.1pt}\discretionary{.}{%
}{.}\hspace{.4pt}1037\discretionary{/}{%
}{/}a0025185}


\bibitem{ggdist}
M.~Kay.
\newblock {\em {ggdist}: {V}isualizations of {D}istributions and {U}ncertainty}.
\newblock Northwestern University, 2023.
\newblock R package version 3.3.0. doi: \textsf{%
10\hspace{.1pt}\discretionary{.}{%
}{.}\hspace{.4pt}5281\discretionary{/}{%
}{/}zenodo\hspace{.1pt}\discretionary{.}{%
}{.}\hspace{.4pt}3879620}


\bibitem{kay2016ish}
M.~Kay, T.~Kola, J.~R. Hullman, and S.~A. Munson.
\newblock When ({I}sh) is {M}y {B}us? {U}ser-{C}entered {V}isualizations of {U}ncertainty in {E}veryday, {M}obile {P}redictive {S}ystems.
\newblock In {\em CHI Conference on Human Factors in Computing Systems}, pp. 5092--5103. Association for Computing Machinery, New York, NY, USA, 2016. doi: \textsf{%
10\hspace{.1pt}\discretionary{.}{%
}{.}\hspace{.4pt}1145\discretionary{/}{%
}{/}2858036\hspace{.1pt}\discretionary{.}{%
}{.}\hspace{.4pt}2858558}


\bibitem{KintschEtAl1975}
W.~Kintsch, E.~Kozminsky, W.~J. Streby, G.~McKoon, and J.~M. Keenan.
\newblock Comprehension and {R}ecall of {T}ext as a {F}unction of {C}ontent {V}ariables.
\newblock {\em Journal of Verbal Learning and Verbal Behavior}, 14(2):196--214, 1975. doi: \textsf{%
10\hspace{.1pt}\discretionary{.}{%
}{.}\hspace{.4pt}1016\discretionary{/}{%
}{/}S0022\discretionary{%
}{-}{-}5371\discretionary{%
}{(}{(}75\discretionary{)}{%
}{)}80065\discretionary{%
}{-}{-}X}


\bibitem{knight:2021}
S.~Knight, N.~Lavan, I.~Torre, and C.~McGettigan.
\newblock {The Influence of Perceived Vocal Traits on Trusting Behaviours in an Economic Game}.
\newblock {\em Quarterly Journal of Experimental Psychology}, 74(10):1747--1754, 2021.
\newblock PMID: 33783278. doi: \textsf{%
10\hspace{.1pt}\discretionary{.}{%
}{.}\hspace{.4pt}1177\discretionary{/}{%
}{/}17470218211010144}


\bibitem{KnowlesLittle2016}
K.~K. Knowles and A.~C. Little.
\newblock {Vocal Fundamental and Formant Frequencies Affect Perceptions of Speaker Cooperativeness}.
\newblock {\em Quarterly Journal of Experimental Psychology}, 69(9):1657--1675, 2016. doi: \textsf{%
10\hspace{.1pt}\discretionary{.}{%
}{.}\hspace{.4pt}1080\discretionary{/}{%
}{/}17470218\hspace{.1pt}\discretionary{.}{%
}{.}\hspace{.4pt}2015\hspace{.1pt}\discretionary{.}{%
}{.}\hspace{.4pt}1091484}


\bibitem{moments}
L.~Komsta and F.~Novomestky.
\newblock {\em Moments, {C}umulants, {S}kewness, {K}urtosis and {R}elated {T}ests}.
\newblock CRAN, 2022.
\newblock R package version 0.14.1.

\bibitem{KurinecWeaver2019}
C.~A. Kurinec and C.~A. Weaver~III.
\newblock {Dialect on Trial: Use of {African American Vernacular English} Influences Juror Appraisals}.
\newblock {\em Psychology, Crime \& Law}, 25(8):803--828, 2019. doi: \textsf{%
10\hspace{.1pt}\discretionary{.}{%
}{.}\hspace{.4pt}1080\discretionary{/}{%
}{/}1068316X\hspace{.1pt}\discretionary{.}{%
}{.}\hspace{.4pt}2019\hspace{.1pt}\discretionary{.}{%
}{.}\hspace{.4pt}1597086}


\bibitem{Lakoff1973}
G.~Lakoff.
\newblock Hedges: {A} {S}tudy in {M}eaning {C}riteria and the {L}ogic of {F}uzzy {C}oncepts.
\newblock {\em Journal of Philosophical Logic}, 2(4):458--508, 1973. doi: \textsf{%
10\hspace{.1pt}\discretionary{.}{%
}{.}\hspace{.4pt}1007\discretionary{/}{%
}{/}BF00262952}


\bibitem{LeeNassBrave2000}
E.~J. Lee, C.~Nass, and S.~Brave.
\newblock {Can Computer-generated Speech Have Gender? An Experimental Test of Gender Stereotype}.
\newblock In {\em CHI'00 Extended Abstracts on Human Factors in Computing Systems}, pp. 289--290, 2000. doi: \textsf{%
10\hspace{.1pt}\discretionary{.}{%
}{.}\hspace{.4pt}1145\discretionary{/}{%
}{/}633292\hspace{.1pt}\discretionary{.}{%
}{.}\hspace{.4pt}633461}


\bibitem{LeeRatanPark2019}
S.~Lee, R.~Ratan, and T.~Park.
\newblock {The Voice Makes the Car: Enhancing Autonomous Vehicle Perceptions and Adoption Intention through Voice Agent Gender and Style}.
\newblock {\em Multimodal Technologies and Interaction}, 3(1):20, 2019. doi: \textsf{%
10\hspace{.1pt}\discretionary{.}{%
}{.}\hspace{.4pt}3390\discretionary{/}{%
}{/}mti3010020}


\bibitem{liu:2016}
L.~Liu, A.~Boone, I.~Ruginski, L.~Padilla, M.~Hegarty, S.~Creem-Regehr, W.~Thompson, C.~Yuksel, and D.~House.
\newblock Uncertainty {V}isualization by {R}epresentative {S}ampling from {P}rediction {E}nsembles.
\newblock {\em IEEE Transactions on Visualization and Computer Graphics}, PP, 09 2016. doi: \textsf{%
10\hspace{.1pt}\discretionary{.}{%
}{.}\hspace{.4pt}1109\discretionary{/}{%
}{/}TVCG\hspace{.1pt}\discretionary{.}{%
}{.}\hspace{.4pt}2016\hspace{.1pt}\discretionary{.}{%
}{.}\hspace{.4pt}2607204}


\bibitem{MehrabianWilliams1969}
A.~Mehrabian and M.~Williams.
\newblock Nonverbal {C}oncomitants of {P}erceived and {I}ntended {P}ersuasiveness.
\newblock {\em Journal of Personality and Social psychology}, 13(1):37--58, 1969. doi: \textsf{%
10\hspace{.1pt}\discretionary{.}{%
}{.}\hspace{.4pt}1037\discretionary{/}{%
}{/}h0027993}


\bibitem{MicrosoftAzureTTS}
{Microsoft Corporation}.
\newblock {Azure Text-to-Speech Service}.
\newblock \url{https://azure.microsoft.com/en-us/services/cognitive-services/text-to-speech/}, 2024.
\newblock Accessed: 2024-06-30.

\bibitem{MicrosoftSpeechService}
{Microsoft Corporation}.
\newblock {Microsoft Speech Service Voice Gallery}, 2024.
\newblock Accessed: 2024-06-30.

\bibitem{MillerEtAl1976}
N.~Miller, G.~Maruyama, R.~J. Beaber, and K.~Valone.
\newblock Speed of {S}peech and {P}ersuasion.
\newblock {\em Journal of Personality and Social Psychology}, 34(4):615--624, 1976. doi: \textsf{%
10\hspace{.1pt}\discretionary{.}{%
}{.}\hspace{.4pt}1037\discretionary{/}{%
}{/}0022\discretionary{%
}{-}{-}3514\hspace{.1pt}\discretionary{.}{%
}{.}\hspace{.4pt}34\hspace{.1pt}\discretionary{.}{%
}{.}\hspace{.4pt}4\hspace{.1pt}\discretionary{.}{%
}{.}\hspace{.4pt}615}


\bibitem{MooshammerEtzrodt2022}
S.~Mooshammer and K.~Etzrodt.
\newblock {Gender Ambiguity in Voice-based Assistants: Gender Perception and Influences of Context}.
\newblock {\em Human-Machine Communication}, 5:49--74, 2022. doi: \textsf{%
10\hspace{.1pt}\discretionary{.}{%
}{.}\hspace{.4pt}3316\discretionary{/}{%
}{/}informit\hspace{.1pt}\discretionary{.}{%
}{.}\hspace{.4pt}869744164360916}


\bibitem{Morgan1990UncertaintyAG}
M.~G. Morgan and M.~Henrion.
\newblock {\em Uncertainty: {A} {G}uide to {D}ealing with {U}ncertainty in {Q}uantitative {R}isk and {P}olicy {A}nalysis}.
\newblock Cambridge University Press, Cambridge, 1990.

\bibitem{Morss2008CommunicatingUI}
R.~Morss, J.~L. Demuth, and J.~K. Lazo.
\newblock Communicating {U}ncertainty in {W}eather {F}orecasts: {A} {S}urvey of the {U.S.} {P}ublic.
\newblock {\em Weather and Forecasting}, 23:974--991, 2008. doi: \textsf{%
10\hspace{.1pt}\discretionary{.}{%
}{.}\hspace{.4pt}1175\discretionary{/}{%
}{/}2008WAF2007088\hspace{.1pt}\discretionary{.}{%
}{.}\hspace{.4pt}1}


\bibitem{nadav2009uncertainty}
L.~Nadav-Greenberg and S.~L. Joslyn.
\newblock Uncertainty {F}orecasts {I}mprove {D}ecision {M}aking {A}mong {N}onexperts.
\newblock {\em Journal of Cognitive Engineering and Decision Making}, 3(3):209--227, 2009. doi: \textsf{%
10\hspace{.1pt}\discretionary{.}{%
}{.}\hspace{.4pt}1518\discretionary{/}{%
}{/}155534309X474460}


\bibitem{OleszkiewiczEtAl2017}
A.~Oleszkiewicz, K.~Pisanski, K.~Lachowicz-Tabaczek, and A.~Sorokowska.
\newblock {Voice-based Assessments of Trustworthiness, Competence, and Warmth in Blind and Sighted Adults}.
\newblock {\em Psychonomic Bulletin \& Review}, 24:856--862, 2017. doi: \textsf{%
10\hspace{.1pt}\discretionary{.}{%
}{.}\hspace{.4pt}3758\discretionary{/}{%
}{/}s13423\discretionary{%
}{-}{-}016\discretionary{%
}{-}{-}1146\discretionary{%
}{-}{-}y}


\bibitem{ottley2019curious}
A.~Ottley, A.~Kaszowska, R.~J. Crouser, and E.~M. Peck.
\newblock {The Curious Case of Combining Text and Visualization}.
\newblock In J.~Johansson, F.~Sadlo, and G.~E. Marai, eds., {\em EuroVis 2019 - Short Papers}. The Eurographics Association, 2019. doi: \textsf{%
10\hspace{.1pt}\discretionary{.}{%
}{.}\hspace{.4pt}2312\discretionary{/}{%
}{/}evs\hspace{.1pt}\discretionary{.}{%
}{.}\hspace{.4pt}20191181}


\bibitem{padillabook:2021}
L.~Padilla, M.~Kay, and J.~Hullman.
\newblock Uncertainty {V}isualization.
\newblock In {\em Wiley StatsRef: Statistics Reference Online}, pp. 1--18. Wiley, 02 2021. doi: \textsf{%
10\hspace{.1pt}\discretionary{.}{%
}{.}\hspace{.4pt}1002\discretionary{/}{%
}{/}9781118445112\hspace{.1pt}\discretionary{.}{%
}{.}\hspace{.4pt}stat08296}


\bibitem{padillaensemble:2017}
L.~Padilla, I.~Ruginski, and S.~Creem-Regehr.
\newblock Effects of {E}nsemble and {S}ummary {D}isplays on {I}nterpretations of {G}eospatial {U}ncertainty {D}ata.
\newblock {\em Cognitive Research: Principles and Implications}, 2, 10 2017. doi: \textsf{%
10\hspace{.1pt}\discretionary{.}{%
}{.}\hspace{.4pt}1186\discretionary{/}{%
}{/}s41235\discretionary{%
}{-}{-}017\discretionary{%
}{-}{-}0076\discretionary{%
}{-}{-}1}


\bibitem{padilla2021uncertain}
L.~M. Padilla, M.~Powell, M.~Kay, and J.~Hullman.
\newblock Uncertain about {U}ncertainty: {H}ow {Q}ualitative {E}xpressions of {F}orecaster {C}onfidence {I}mpact {D}ecision-making with {U}ncertainty {V}isualizations.
\newblock {\em Frontiers in Psychology}, 11:579267, 2021. doi: \textsf{%
10\hspace{.1pt}\discretionary{.}{%
}{.}\hspace{.4pt}3389\discretionary{/}{%
}{/}fpsyg\hspace{.1pt}\discretionary{.}{%
}{.}\hspace{.4pt}2020\hspace{.1pt}\discretionary{.}{%
}{.}\hspace{.4pt}579267}


\bibitem{palan2018prolific}
S.~Palan and C.~Schitter.
\newblock {Prolific. ac—A Subject Pool for Online Experiments}.
\newblock {\em Journal of Behavioral and Experimental Finance}, 17:22--27, 2018. doi: \textsf{%
10\hspace{.1pt}\discretionary{.}{%
}{.}\hspace{.4pt}1016\discretionary{/}{%
}{/}j\hspace{.1pt}\discretionary{.}{%
}{.}\hspace{.4pt}jbef\hspace{.1pt}\discretionary{.}{%
}{.}\hspace{.4pt}2017\hspace{.1pt}\discretionary{.}{%
}{.}\hspace{.4pt}12\hspace{.1pt}\discretionary{.}{%
}{.}\hspace{.4pt}004}


\bibitem{pandey2023you}
S.~Pandey, O.~G. McKinley, R.~J. Crouser, and A.~Ottley.
\newblock {Do You Trust What You See? Toward A Multidimensional Measure of Trust in Visualization}.
\newblock {\em IEEE}, pp. 26--30, 2023. doi: \textsf{%
10\hspace{.1pt}\discretionary{.}{%
}{.}\hspace{.4pt}1109\discretionary{/}{%
}{/}VIS54172\hspace{.1pt}\discretionary{.}{%
}{.}\hspace{.4pt}2023\hspace{.1pt}\discretionary{.}{%
}{.}\hspace{.4pt}00014}


\bibitem{pias2024impact}
S.~B.~H. Pias, R.~Huang, D.~Williamson, M.~Kim, and A.~Kapadia.
\newblock {The Impact of Perceived Tone, Age, and Gender on Voice Assistant Persuasiveness in the Context of Product Recommendations}.
\newblock {\em arXiv preprint arXiv:2405.04791}, 2024.

\bibitem{qualtrics}
Qualtrics, Provo, Utah, USA.
\newblock {\em Qualtrics}, 2023.
\newblock Version: June 2024.

\bibitem{rComputing}
{R Core Team}.
\newblock {\em R: A Language and Environment for Statistical Computing}.
\newblock R Foundation for Statistical Computing, Vienna, Austria, 2023.

\bibitem{sacha:2016}
D.~Sacha, H.~Senaratne, B.~C. Kwon, G.~Ellis, and D.~A. Keim.
\newblock The {R}ole of {U}ncertainty, {A}wareness, and {T}rust in {V}isual {A}nalytics.
\newblock {\em IEEE Transactions on Visualization and Computer Graphics}, 22(1):240--249, 2016. doi: \textsf{%
10\hspace{.1pt}\discretionary{.}{%
}{.}\hspace{.4pt}1109\discretionary{/}{%
}{/}TVCG\hspace{.1pt}\discretionary{.}{%
}{.}\hspace{.4pt}2015\hspace{.1pt}\discretionary{.}{%
}{.}\hspace{.4pt}2467591}


\bibitem{SchererEtAl1973}
K.~R. Scherer, H.~London, and J.~J. Wolf.
\newblock The {V}oice of {C}onfidence: {P}aralinguistic {C}ues and {A}udience {E}valuation.
\newblock {\em Journal of Research in Personality}, 7:31--44, 1973. doi: \textsf{%
10\hspace{.1pt}\discretionary{.}{%
}{.}\hspace{.4pt}1016\discretionary{/}{%
}{/}0092\discretionary{%
}{-}{-}6566\discretionary{%
}{(}{(}73\discretionary{)}{%
}{)}90030\discretionary{%
}{-}{-}5}


\bibitem{Spiegelhalter:20111}
D.~Spiegelhalter, M.~Pearson, and I.~Short.
\newblock Visualizing {U}ncertainty {A}bout the {F}uture.
\newblock {\em Science}, 333(6048):1393--1400, 2011. doi: \textsf{%
10\hspace{.1pt}\discretionary{.}{%
}{.}\hspace{.4pt}1126\discretionary{/}{%
}{/}science\hspace{.1pt}\discretionary{.}{%
}{.}\hspace{.4pt}1191181}


\bibitem{StentSyrdalMishra2011}
A.~Stent, A.~Syrdal, and T.~Mishra.
\newblock {On the Intelligibility of Fast Synthesized Speech for Individuals with Early-onset Blindness}.
\newblock In {\em Proceedings of the 13th International ACM SIGACCESS Conference on Computers and Accessibility}, pp. 211--218, 2011. doi: \textsf{%
10\hspace{.1pt}\discretionary{.}{%
}{.}\hspace{.4pt}1145\discretionary{/}{%
}{/}2049536\hspace{.1pt}\discretionary{.}{%
}{.}\hspace{.4pt}2049574}


\bibitem{stokes2024delays}
C.~Stokes, C.~Sanker, B.~Cogley, and V.~Setlur.
\newblock {From Delays to Densities: Exploring Data Uncertainty through Speech, Text, and Visualization}.
\newblock In {\em Computer Graphics Forum}, vol.~43, p. e15100. Wiley Online Library, 2024. doi: \textsf{%
10\hspace{.1pt}\discretionary{.}{%
}{.}\hspace{.4pt}1111\discretionary{/}{%
}{/}cgf\hspace{.1pt}\discretionary{.}{%
}{.}\hspace{.4pt}15100}


\bibitem{stokes2024mixing}
C.~Stokes, C.~Sanker, B.~Cogley, and V.~Setlur.
\newblock {Mixing Modes: Active and Passive Integration of Speech, Text, and Visualization for Communicating Data Uncertainty}.
\newblock In {\em Eurographics Conference on Visualization (EuroVis)}. The Eurographics Association, 2024.

\bibitem{Sundar2000}
S.~S. Sundar.
\newblock Multimedia {E}ffects on {P}rocessing and {P}erception of {O}nline {N}ews: A {S}tudy of {P}icture, {A}udio, and {V}ideo {D}ownloads.
\newblock {\em Journalism \& Mass Communication Quarterly}, 77(3):480--499, 2000. doi: \textsf{%
10\hspace{.1pt}\discretionary{.}{%
}{.}\hspace{.4pt}1177\discretionary{/}{%
}{/}107769900007700302}


\bibitem{SzarvasEtAl2012}
G.~Szarvas, V.~Vincze, R.~Farkas, G.~M{\'o}ra, and I.~Gurevych.
\newblock {Cross-genre and Cross-domain Detection of Semantic Uncertainty}.
\newblock {\em Computational Linguistics}, 38(2):335--367, 2012. doi: \textsf{%
10\hspace{.1pt}\discretionary{.}{%
}{.}\hspace{.4pt}1162\discretionary{/}{%
}{/}COLI\_a\_00098}


\bibitem{Thomson2005ATF}
J.~R. Thomson, E.~G. Hetzler, A.~M. MacEachren, M.~Gahegan, and M.~Pavel.
\newblock A {T}ypology for {V}isualizing {U}ncertainty.
\newblock In R.~F. Erbacher, J.~C. Roberts, M.~T. Grohn, and K.~Borner, eds., {\em Proceedings of the Society of Photo-Optical Instrumentation Engineers (SPIE) 5669}, pp. 146--157. SPIE, 2005. doi: \textsf{%
10\hspace{.1pt}\discretionary{.}{%
}{.}\hspace{.4pt}1117\discretionary{/}{%
}{/}12\hspace{.1pt}\discretionary{.}{%
}{.}\hspace{.4pt}587254}


\bibitem{Thorne2019TheTO}
N.~Thorne, A.~K.-T. Yip, W.~P. Bouman, E.~Marshall, and J.~Arcelus.
\newblock {The Terminology of Identities Between, Outside and Beyond the Gender Binary – A Systematic Review}.
\newblock {\em International Journal of Transgender Health}, 20:138 -- 154, 2019.

\bibitem{TolmeijerEtAl2021}
S.~Tolmeijer, N.~Zierau, A.~Janson, J.~S. Wahdatehagh, J.~M.~M. Leimeister, and A.~Bernstein.
\newblock {Female by Default? Exploring the Effect of Voice Assistant Gender and Pitch on Trait and Trust Attribution}.
\newblock In {\em CHI Conference on Human Factors in Computing Systems}, 2021. doi: \textsf{%
10\hspace{.1pt}\discretionary{.}{%
}{.}\hspace{.4pt}1145\discretionary{/}{%
}{/}3411763\hspace{.1pt}\discretionary{.}{%
}{.}\hspace{.4pt}3451623}


\bibitem{tsai2008effects}
C.~I. Tsai, J.~Klayman, and R.~Hastie.
\newblock Effects of {A}mount of {I}nformation on {J}udgment {A}ccuracy and {C}onfidence.
\newblock {\em Organizational Behavior and Human Decision Processes}, 107(2):97--105, 2008. doi: \textsf{%
10\hspace{.1pt}\discretionary{.}{%
}{.}\hspace{.4pt}1016\discretionary{/}{%
}{/}j\hspace{.1pt}\discretionary{.}{%
}{.}\hspace{.4pt}obhdp\hspace{.1pt}\discretionary{.}{%
}{.}\hspace{.4pt}2008\hspace{.1pt}\discretionary{.}{%
}{.}\hspace{.4pt}01\hspace{.1pt}\discretionary{.}{%
}{.}\hspace{.4pt}005}


\bibitem{tversky1974judgment}
A.~Tversky and D.~Kahneman.
\newblock Judgment under {U}ncertainty: {H}euristics and {B}iases: {B}iases in {J}udgments {R}eveal {S}ome {H}euristics of {T}hinking under {U}ncertainty.
\newblock {\em Science}, 185(4157):1124--1131, 1974. doi: \textsf{%
10\hspace{.1pt}\discretionary{.}{%
}{.}\hspace{.4pt}1126\discretionary{/}{%
}{/}science\hspace{.1pt}\discretionary{.}{%
}{.}\hspace{.4pt}185\hspace{.1pt}\discretionary{.}{%
}{.}\hspace{.4pt}4157\hspace{.1pt}\discretionary{.}{%
}{.}\hspace{.4pt}1124}


\bibitem{WaytzHeafnerEpley2014}
A.~Waytz, J.~Heafner, and N.~Epley.
\newblock The {M}ind in the {M}achine: {A}nthropomorphism {I}ncreases {T}rust in an {A}utonomous {V}ehicle.
\newblock {\em Journal of Experimental Social Psychology}, 52:113--117, 2014. doi: \textsf{%
10\hspace{.1pt}\discretionary{.}{%
}{.}\hspace{.4pt}1016\discretionary{/}{%
}{/}j\hspace{.1pt}\discretionary{.}{%
}{.}\hspace{.4pt}jesp\hspace{.1pt}\discretionary{.}{%
}{.}\hspace{.4pt}2014\hspace{.1pt}\discretionary{.}{%
}{.}\hspace{.4pt}01\hspace{.1pt}\discretionary{.}{%
}{.}\hspace{.4pt}005}


\bibitem{W3CSSML}
{World Wide Web Consortium (W3C)}.
\newblock {Speech Synthesis Markup Language (SSML) Version 1.0}.
\newblock \url{https://www.w3.org/TR/speech-synthesis/}, 2004.
\newblock Accessed: 2024-06-30.

\bibitem{xiong2019examining}
C.~Xiong, L.~Padilla, K.~Grayson, and S.~Franconeri.
\newblock Examining the {C}omponents of {T}rust in {M}ap-based {V}isualizations.
\newblock In {\em 1st EuroVis Workshop on Trustworthy Visualization, TrustVis 2019}, pp. 19--23. The Eurographics Association, 2019. doi: \textsf{%
10\hspace{.1pt}\discretionary{.}{%
}{.}\hspace{.4pt}2312\discretionary{/}{%
}{/}trvis\hspace{.1pt}\discretionary{.}{%
}{.}\hspace{.4pt}20191186}


\end{thebibliography}
\end{document}